\documentclass[sigconf]{acmart}

\AtBeginDocument{%
  }

\setcopyright{none}
\copyrightyear{2023}
\acmYear{2023}
\setcopyright{acmlicensed}

\acmConference[CSCW '23 Companion] {Computer Supported Cooperative Work and Social Computing}{October 14--18, 2023}{Minneapolis, MN, USA}

\acmBooktitle{Computer Supported Cooperative Work and Social Computing (CSCW '23 Companion), {October 14--18, 2023, Minneapolis, MN, USA}}

\acmPrice{15.00}

\acmDOI{10.1145/XXXXXX.XXXXXX}

\acmISBN{979-8-4007-0129-0/23/10}

\newcommand\footnotetextcopyrightpermission[1]{} %
\usepackage{amsmath}
\usepackage{titlesec}
\usepackage{pdfpages}
\usepackage{setspace}
\usepackage{cite}
\usepackage{epsfig}
\usepackage{wrapfig}

\usepackage{amsfonts}
\usepackage{verbatim}
\usepackage{xspace}
\usepackage{xcolor}

\usepackage{wrapfig}

\usepackage{lipsum}
\usepackage{ragged2e}
\usepackage[inline]{enumitem}
\usepackage{tocloft}
\usepackage{etoolbox}

\usepackage{dirtytalk}
\usepackage{booktabs}

\usepackage{amsfonts}       
\usepackage{nicefrac}       
\usepackage{microtype}      
\usepackage{empheq}
\usepackage{fancyhdr}
\usepackage{amsmath}
\usepackage{hyperref}

\pdfoutput=1
\usepackage{multirow}
\usepackage{colortbl}

\usepackage{xcolor}
\usepackage{pict2e}
\usepackage{comment}

\usepackage{enumitem}
\usepackage{placeins}

\usepackage{graphicx}

\begin{document}


\newcommand*{\img}[1]{%
    \raisebox{-.0\baselineskip}{%
        \includegraphics[
        height=\baselineskip,
        width=\baselineskip,
        keepaspectratio,
        ]{#1}%
    }%
}

\title[\system: Delivering Heterogeneous AI Explanations via Conversations to Support Human-AI Scientific Writing]{\system
\hspace*{-0.08in}
\img{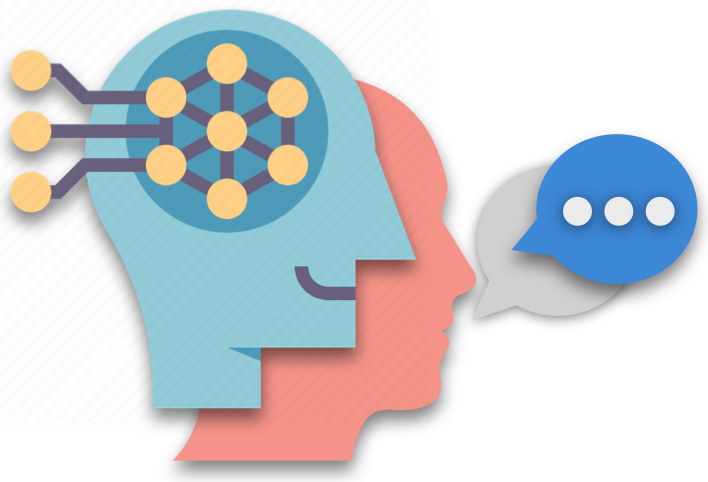}: Delivering Heterogeneous AI Explanations via Conversations to Support Human-AI Scientific Writing}





\author{Hua Shen}
\email{huashen@umich.edu}
\affiliation{%
    \institution{University of Michigan}
    \institution{Pennsylvania State University}
  \country{USA}}

\author{Chieh-Yang Huang}
\email{chiehyang@psu.edu}
\affiliation{%
  \institution{Pennsylvania State University}
    \country{}
  }

\author{Tongshuang Wu}
\email{sherryw@cs.cmu.edu}
\affiliation{%
  \institution{Carnegie Mellon University}
    \country{}
  }

\author{Ting-Hao (Kenneth) Huang}
\email{txh710@psu.edu}
\affiliation{%
  \institution{Pennsylvania State University}
\country{}
  }

\renewcommand{\shortauthors}{Hua Shen, Chieh-Yang Huang, Tongshuang Wu, Kenneth Huang.}

\newcommand*{\numimg}[1]{%
    \raisebox{-.15\baselineskip}{%
        \includegraphics[
        height=.9\baselineskip,
        width=.9\baselineskip,
        ]{#1}%
    }%
}

\newcommand{\hua}[1]{{\small\textcolor{orange}{\bf [#1 --Hua]}}}

\newcommand{\system}{\textsl{\texttt{ConvXAI}}\xspace}
\newcommand{\baseline}{\textsl{\texttt{SelectXAI}}\xspace}

\newcommand{\convxai}{{conversational XAI \xspace}}

\newcommand{\Convxai}{\textsl{\textsc{Conversational XAI}}\xspace}

\newcommand{\done}[1]{{\textcolor{blue}{[Done]#1}}}
\newcommand{\eg}{\emph{e.g.,}\xspace}
\newcommand{\ie}{\emph{i.e.,}\xspace}


\definecolor{principle}{HTML}{000000}
\definecolor{section}{HTML}{346CF0}
\definecolor{figure}{HTML}{346CF0}

\newcommand{\Newdot}{{\leavevmode\put(0,0){\color{red}\circle*{2.5}}}}

\begin{abstract}

While various methods of AI explanation (XAI) have been proposed to interpret AI systems, users still face challenges in obtaining the information they require. Previous research has suggested the use of chatbots to cater to human needs dynamically, yet there is limited exploration of how conversational XAI agents can be effectively designed for practical use. This paper focuses on applying Conversational XAI to AI-assisted human scientific writing tasks. Drawing inspiration from human linguistics and formative studies with 7 users of diverse backgrounds, we identify four key design rationales for practically useful Conversational XAI: addressing diverse user questions (``multifaceted''), providing details on-demand (``controllability''), proactively tutoring  XAI suggestions (``mix-initiative''), and tracking dialog history for contexts (``context-aware drill-down''). These rationales are implemented in an interactive prototype called \system\footnote{See the \system system code at: \url{https://github.com/huashen218/convxai.git}.}, which facilitates AI-assisted scientific writing interaction with heterogeneous AI explanations through a dialogue interface\footnote{See the \system unified XAI API at: \url{https://github.com/huashen218/convxai/blob/main/notebook_unified_XAI_API/convxai_unified_api.ipynb}.}.
Through two within-subjects studies with 21 users, we demonstrate that \system is more useful, compared with a GUI-based baseline prototype, for humans in perceiving the understanding and writing improvement, and improving the writing process in terms of productivity and sentence quality. The paper concludes by discussing the limitations of \system and proposing potential avenues for future research in useful XAI with conversations or interactions.

\end{abstract}


%
%


%


\begin{teaserfigure}
  \includegraphics[width=.98\textwidth]{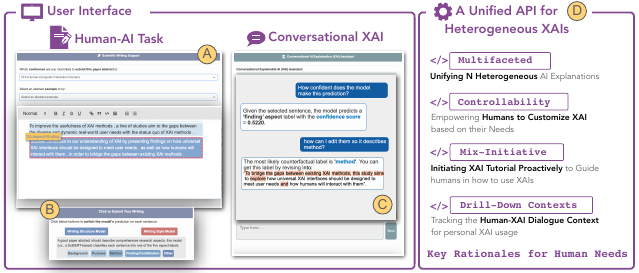}
  \caption{
    An overview of \system to support human-AI scientific writing with heterogeneous AI explanations via dialog.
   \system includes a front-end User Interface to~\hspace*{.0em}\numimg{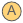}\hspace*{-.1em} support human-AI collaborative task interaction, \hspace*{.0em}\numimg{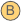}\hspace*{-.1em} check AI models and predictions, and \hspace*{.0em}\numimg{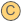}\hspace*{-.1em} inquire about heterogeneous AI explanations via dialogue. Also, \system involves a back-end deep learning server to generate AI predictions and explanations, which is embedded with \hspace*{.0em}\numimg{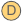}\hspace*{-.1em} a unified API for generating heterogeneous AI explanations that are designed to cater to practical human use needs.
  }
  \label{fig:teaser}
\end{teaserfigure}

\maketitle

\section{Introduction}
\label{sec:introduction}

%
%
The advancement of deep learning has led to breakthroughs in a number of artificial intelligence systems (AI).
Yet, the superior performance of AI systems is often achieved at the expense of the interpretability of deep learning models~\citep{miller2019explanation}.
%
%
%
To address this challenge, researchers have developed a collection of eXplainable AI (XAI) methods that aim to enhance human understanding of AI from various perspectives~\citep{shen2020useful,shen2021explaining}. These methods typically focus on answering specific XAI questions of interest to users. For example, saliency maps and feature attributions~\citep{lime,shap} highlight key rationales behind AI predictions to address "why" questions, while counterfactual explanations perturb input to explore "why X not Y" scenarios that affect model behavior~\citep{miller2019explanation,wu2021polyjuice}.
Despite their potential, 
the usefulness of XAI methods in real-world applications has yielded inconsistent findings~\citep{bansal2021does,PoursabziSangdeh2021ManipulatingAM,shen2020useful}. While some studies demonstrate that different explanations can support specific use cases, such as model debugging~\citep{lertvittayakumjorn2021explanation} and human-AI collaboration~\citep{gonzalez2020human}, others reveal limitations in enhancing users' ability to simulate model predictions~\citep{shen-etal-2022-shortest} or understand AI errors~\citep{shen2020useful}. To bridge this gap, researchers have explored the mismatch between real-world user demands and existing XAI methods. \citet{shen2021explaining}, for instance, compare practical user questions~\citep{liao2020questioning} with over 200 XAI studies and identify a bias in current methods towards certain types of XAI questions, neglecting others. 
Additionally, users also tend to have \emph{multiple}, \emph{dynamic} and sometimes \emph{interdependent} questions on AI explanations~\citep{lakkaraju2022rethinking,tsai2021exploring,}.

Addressing this array of questions necessitates an integration of heterogeneous AI explanations. Taking inspiration from the flexibility of dialog systems~\citep{jurafsky2000speech,fast2018iris}, prior work has envisioned the concept of "explainability as a dialogue" to accommodate diverse user needs and mitigate cognitive load~\citep{slack2022talktomodel,lakkaraju2022rethinking}. For instance, \citet{lakkaraju2022rethinking} discovered that decision-makers strongly prefer interactive explanations in the form of natural language dialogue. However, there is a dearth of exploration regarding the design of conversational XAI systems to meet practical user needs and understand user reactions.

In this paper, we investigate the potential of conversational XAI in the context of practical human-AI collaborative writing. Through formative user studies on a preliminary system and a review of human conversation characteristics, we identify four design rationales for conversational XAI: addressing various user questions (``multi-faceted''), actively suggesting and accepting follow-up questions (``mix-initiative'' and ``context-aware drill-down''), and providing on-demand details (``controllability''). Guided by these rationales, we develop a conversational XAI prototype system called \system, which incorporates the four user-oriented XAI principles. Moreover, we evaluate the potential of ConvXAI in the realm of human-AI scientific writing, where writers leverage ConvXAI to improve their paper abstracts for submission to top-tier research conferences. In this use case, \system assists users in interacting with two AI writing models that assess the structure and quality of abstracts at the sentence level. 
Users can engage in dialogue with \system to comprehend the writing feedback and enhance their papers with the aid of heterogeneous AI explanations.

We conducted two within-subject user studies to evaluate the \system system. We compared \system with \baseline, a traditional GUI-based universal XAI system that displays all XAIs on the interface in a collapsible manner (Figure~\ref{fig:baseline}). In the first user study, involving an open-ended writing task with 13 participants, we found that the majority of users perceived \system to be more useful in understanding AI writing feedback and improving their own writing. These results further confirmed the reduced cognitive load and effectiveness of the four user-oriented design principles. Additionally, in the second user study, which focused on a well-defined writing task with 8 rejoining participants, we collected the users' writing artifacts generated using both \system and \baseline systems. We evaluated these artifacts using both human evaluators and auto-metrics. The analysis revealed that both \system and \baseline assisted users in producing better writing based on the built-in auto-metrics, with \system proving particularly valuable for improving writing quality. However, we observed a misalignment between the measurements of the human evaluator and the auto-metrics, indicating the importance of designing AI model predictions to align with human expectations.
Building upon these studies and findings, we further contribute insights into the practical human usage patterns of XAI in \system and core ingredients of useful XAI systems for future XAI work. We conclude this work by discussing its limitations and outlining future research directions.

\section{Related Work}
\label{sec:literature}

\subsection{Human-Centered AI Explanations}

Earlier studies in the fields of Explainable Artificial Intelligence (XAI) primarily focus on developing different XAI techniques, which aims to explain \emph{why the model arrives at the predictions.} This line of studies can be broadly categorized into generating post-hoc interpretations for well-trained deep learning models~\citep{geval:2019:blackboxnlp} and designing self-explaining models~\citep{shen-etal-2022-shortest,lei-etal-2016-rationalizing,shi2016does}. 
In specific, the majority of XAI methods aim to provide post-hoc interpretations either for each input instance ({\ie} named ``local explanations'')~
\citep{de-cao-etal-2020-decisions,kim-etal-2020-interpretation,sankar-etal-2019-neural} or for providing a global view of how the AI model works ({\ie} named ``global explanations'')~\citep{ribeiro2018anchors}, where our study covers both of them. 
Additionally, XAI approaches are also divided into different formats~\citep{shen2021explaining},
including example-based~\citep{feng2019can}, feature-based~\citep{lime}, free text-based~\citep{rajani-etal-2019-explain,NEURIPS2018_4c7a167b}, rule-based explanations~\citep{ribeiro2018anchors}, etc, where our study covers a range of XAI formats.

Despite the increasing number of XAI approaches have been proposed, evaluating AI with humans is still a challenging problem.
Doshi-Velez and Kim~\citep{doshi:2017:towards} propose a taxonomy of interpretability evaluation including ``application-grounded'', ``human-grounded'' and ``functionally-grounded'' evaluation metrics based on different levels of human involvement and application tasks.
The majority of the proposed XAI approaches are commonly validated effectively using the ``functionally-grounded'' evaluation methods~\citep{jacovi2020towards,promisperil,wiegreffe-pinter-2019-attention}, 
which seek for automatic metrics ({\eg} ``plausibility'') on proxy tasks without real human participations~\citep{mohankumar2020towards,bastings-filippova-2020-elephant,10.1145/3359158}.

Furthermore, we can see burgeoning efforts being put around involving real humans in evaluating AI explanations under the theme of ``human-centered explainable AI''. The state-of-the-art XAI methods are applied to real human tasks, such as assessing human understanding~\citep{shen2020useful}, human simulatability~\citep{ribeiro2018anchors,shen-etal-2022-shortest}, human trust and satisfaction on AI predictions~\citep{smith2020no,reasonchains:2019:emnlp}, and human-AI teamwork performance~\citep{chu2020visual}, etc~\citep{hase2020evaluating,feng2019can,ghai2020explainable}. 
However, many human studies show that AI explanations are not always helpful for human understanding in tasks such as simulating model prediction~\citep{shen-etal-2022-shortest}, analyzing model failures~\citep{shen2020useful}, human-AI team collaboration~\citep{Bansal2021DoesTW}. For instance, \citet{Bansal2021DoesTW} conducted human studies to investigate if XAI helps achieve complementary team performance and showed that none of the explanation conditions produced an accuracy significantly higher than the simple baseline of showing confidence.

In response, a line of work dives deep into the gaps between real-world user demands and the status quo XAI methods. Their findings reveal that users tend to ask \emph{multiple}, \emph{dynamic}, and sometimes \emph{interdependent} questions on AI explanations, whereas state-of-the-art XAI methods are mostly unable to satisfy.
Although GUI-based XAI systems, which integrate multiple XAI into one interface, can potentially mitigate this issue, they inevitably suffer from the drawbacks, such as cognitive overload, frequent UI updates, etc.

Therefore, prior studies envision the potential of ``Explainability as a Dialogue'' to balance the cognitive load with the diverse user needs~\citep{slack2022talktomodel,lakkaraju2022rethinking,marrakchi2021explaining,tsai2021exploring,sun2022exploring}. 
For example, through interviews with healthcare professionals and policymakers,  \citet{lakkaraju2022rethinking} found that decision-makers strongly prefer interactive explanations with natural language dialogue forms and thereby advocated for interactive explanations. 
Nevertheless, there has been little exploration of how a conversational XAI system should be designed in practice and how users might react to it.
Our studies aim to resolve this problem by incorporating practical user needs into the 
conversational XAI design, propose a user-oriented conversational universal XAI interface and investigate human behaviors during using these systems.

\subsection{Conversational AI Systems}

Our work is situated within the rich body of conversational AI or chatbots studies, which entails a long research history in the NLP~\citep{coqa,li2019acute} and HCI fields~\citep{setlur2022you,fast2018iris}. Jurafsky~\citep{jurafsky2000speech} proposes that conversation between humans is an intricate and complex join activity, which entails a set of imperative properties: {\em multiple turns}, {\em common grounding}; {\em dialogue structure},  {\em mixed-initiative}.
By incorporating these properties, conversational interactions are also shown to significantly contribute to establishing long-term rapport and trust between humans and systems~\citep{bickmore2001relational}. 
User interaction experience can be improved by a set of factors from the conversational AI systems~\citep{setlur2022you}. For example, Chaves and Gerosa~\citep{choudhry2020once} describe how human-like social characteristics, such as conversational intelligence and manners, may benefit the user experience.

These principles and theories inform us to design a conversational AI explanation system that fulfills the diverse user needs in practice. Our study is deeply rooted in the conversational explanations in XAI -- the users request their demanded explanations through the chatbot-based AI assistants~\citep{sokol2018glass, tsai2021exploring}. Previous studies have explored the effectiveness of interactive dialogues in explaining online symptom checkers (OSCs)~\citep{tsai2021exploring, sun2022exploring}. For example, Tsai et al.~\citep{tsai2021exploring} intervened in the diagnostic and triage recommendations of the OSCs with three types of explanations ({\ie rationale-based, feature-based and example-based explanations}) during the conversational flows.
The findings yield four implications for future OSC designs, which include empowering users with more control, generating multifaceted and context-aware explanations, and being cautious of the potential downsides.

However, these existing conversational AI explanation systems are still in the preliminary stage, which only provides one type of explanation and disables users from selecting different explanation types.
Also, these are far from being able to incorporate user feedback into producing AI explanations ({\eg enable users to choose counterfactual prediction foil}) and produce personalized explanations for users' individual needs.
In addition, these conversational AI explanation systems are primarily applied to improve system transparency and comprehensibility, thus helping users understand and build trust in the systems.
Little attention has been paid to examining {\em if} and {\em how} conversational AI explanations can be indeed useful for users to improve their performance in human-AI collaborative tasks.
%
%

Our work improves the conversational AI explanation systems from two perspectives: i) we focus on AI tasks where the human's goal is to improve their task performance ({\ie} scientific writing) rather than merely gain an understanding of the AI predictions; ii) we identify four design principles and incorporate them into the empirical system design for further evaluation with human tasks.
Our work aims to further unleash the capability of conversational AI explanations and make them more useful for human tasks.

\begin{table*}
  \includegraphics[width=\textwidth]{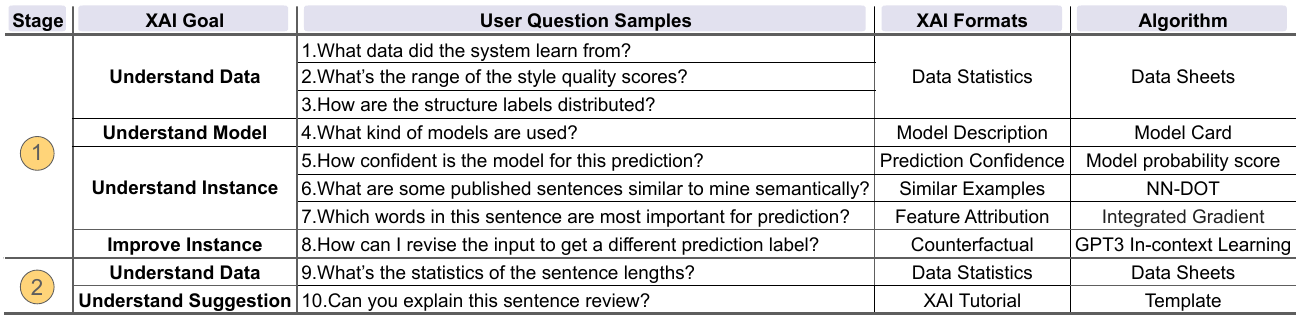}
  \caption{\system covers ten types of user questions (\emph{i.e.,} Data Statistic, Model Description, Feature Attribution, etc.) serving to five different XAI goals  (\emph{e.g.,} Understand Model, Understand Data, Improve Instance, etc.). Stage (1) shows eight XAIs included in the formative study, and Stage (2) indicates two added XAIs in \system.
  }
  \label{fig:xai_questions}. 
\end{table*}


\subsection{AI Writing Support Tools}

The improvements in large language models (LMs) like GPT3~\citep{gpt3} and Meena~\citep{meena} have provided unprecedented language generation power. This leads to an increasing interest in how these new technologies may support writers with AI-assisted writing support tools~\citep{lee2022coauthor}. 
In these human-AI collaborative writing tasks, the writers interact with AI writing support tools not only for understanding its assessment but also aim to leverage its feedback to improve the human writing output~\citep{huang2020heteroglossia}. 
%
A few technologies are developed to support human writing. Many of them focused on \emph{lower-level linguistic improvement}, such as proofreading, text generation, grammar correction, auto-completion, etc. 
For instance, Roemmele and Gordon~\citep{roemmele2015creative} proposed a Creative Help system that uses a recurrent neural network model to generate suggestions for the next sentence. 
Furthermore, a few studies propose AI assistants that leverage the generation capability of the language models to \emph{generate inspirations} to assist the writers' ideation process~\citep{gero2022design,wang2022interpretable,coenen2021wordcraft}. 
For instance, 
Wordcraft~\citep{coenen2021wordcraft} is an AI-assisted editor proposed for story writing, in which a writer and a dialogue system collaborate to write a story. The system further supports natural language generation to users including planning, writing and editing the story creation.

In addition, there are a number of studies that design AI assistants to \emph{provide assessment and feedback} to help improve human writings iteratively~\citep{du2022read,shen2023parachute}. 
For example, Huang et al.~\citep{huang2018feedback} argue that writing, as a complex creative task, demands rich feedback in the writing revision process. They present Feedback Orchestration to guide writers to integrate feedback into revisions by a rhetorical structure. 
More studies are proposed for AI-assisted peer review~\citep{checco2021ai}. For example, Yuan et al.~\citep{yuan2021can} automate the scientific review process that uses LLMs to generate reviews for scientific papers.

In this work, we apply conversational AI explanations to human-AI scientific writing tasks, in which humans submit their writings to the system and iteratively make a sequence of small decision-making processes based on AI feedback and explanations.
As \emph{writing is a goal-directed thinking process}~\citep{gero2022design}. 
The goal of the \convxai system is to support writers to \emph{understand the feedback} and further \emph{improve their writing} outputs. Therefore, we aim to evaluate the effects of conversational AI explanations in terms of not only helping users understand the AI prediction but also improving writing performance.

\section{Understanding Practical User Demands in Conversational XAI}
\label{sec:formative}

\begin{figure*}[!t]
  \includegraphics[width=0.85\textwidth]{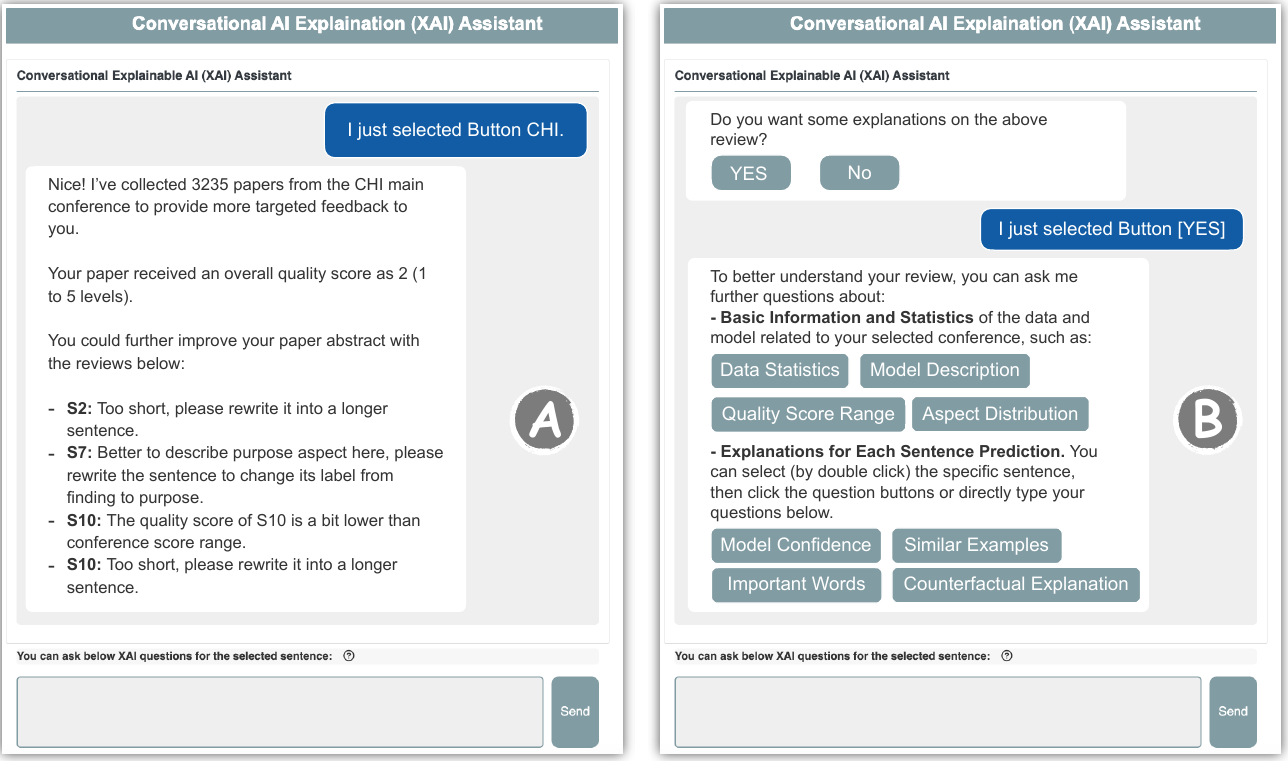}
  \caption{An overview of User Interface (UI) for the pilot study. (A) shows the recommended edits from the writing models, and (B) displays a range of XAI buttons for users to choose from for viewing AI explanations.
  }
  \label{fig:pilot}
\end{figure*}


Due to the unique characteristics of AI-assisted human scientific writing tasks and the early status of conversational XAI systems, we see a lack of established designs and techniques of conversational AI explanations that can cater to user needs in scientific writing support tasks. 
Therefore, we first analyze the practical user demands of conversational XAI by conjecturing a system walkthrough in a usage scenario with a student submitting her CHI paper (Section~\ref{sec:user-scenario}), and then conducting a formative study with seven users of diverse backgrounds (Section~\ref{sec:formative}). We summarize the resulting four design rationales in Section~\ref{sec:principles}. 


\subsection{Example User Scenario}
\label{sec:user-scenario}


Gloria is a Ph.D. student in the CHI research field. While she has already finished a paper draft, she wants to use the system to receive more paper review feedback on her paper abstract writing, so that the paper would get a higher chance of being accepted by the CHI conference.
She is especially curious about 
\emph{What would be the review feedback of my paper abstract}? \emph{Why would the system give me this feedback}? \emph{How should I improve my writing to get a better paper abstract}? To answer these questions, Gloria starts to interact with the system with these questions in mind.
First, she is asked to {choose the target conference} 
she wants to submit her paper abstract.
%
%
After choosing CHI as the target conference, Gloria can see the abstract example options and writing editor panel, so that Gloria can \emph{edit her abstract content} and then submit her abstract to get AI assistant assessments on each sentence.

For example, one piece of writing review that Gloria received is ``Sentence 3: Based on the sentence labels' percentage and order in your abstract, it is suggested to write your \emph{background} at this sentence, rather than describing \emph{purpose} here.''.
Before diving deeper into understanding the predictions, Gloria first wants to assess if she should trust the models by understanding how the model and data work in this \system system. So she asks \emph{``What data did the system use?''} and \emph{``What kind of models is the \system using?''}. After learning that the \system is using the state-of-the-art language models and the data is the collection of the latest five years from CHI, Gloria decides to trust the system and proceed with the AI explanations.
At the next stage, Gloria is wondering \emph{why the system suggests she describe the background instead of purpose in the sentence 3}. By asking {\emph{``What words make the assistant think it is describing the purpose?''}}, she learns that the ``purpose'' aspect prediction is attributed to the top 6 important words, including ``examine'', ``paper'', ``conversational'', ``xai'', ``scientific'', ``writing'' (feature attribution explanation).
Furthermore, Gloria wants to know {\emph{``how can I edit them so it describes background''}} and is suggested to remove the ``in this paper, we examine'' words at the beginning and add ``is yet to be explored'' in the end.
(counterfactual explanation)
After interacting with the XAI agent with multi-turn dialogues, she \emph{understands the system,  predictions and reviews} better.
Finally, Gloria \emph{revises the sentence} based on her understanding with the help of XAI agent and \emph{re-submit the abstract}. The structure review is successfully resolved.
Gloria can then move on to the next sentence.

\subsection{Formative Study}
\label{sec:formative}


In the early phase of the project, we conducted a formative study to inform ourselves about how humans leverage AI explanations to achieve their AI-assisted scientific writing tasks, and the common limitations and needs necessary for enhancing human performance. This is primarily to help us develop a set of design rationales listed in Section~\ref{sec:principles} to motivate system designs.

%
%
%
%

\subsubsection{{AI Tasks and AI Explanation Design.}}
To form the human-AI interactive writing scenario, we develop two AI writing models to generate writing structure and style predictions, respectively. The writing structure model gives each sentence a research aspect label, indicating which aspect the sentence is describing among the five categories (\emph{i.e.,} background, purpose, method, contribution/finding, and others).
On the other hand, the writing style model provides each sentence a style quality score assessing ``how well the writing style of this sentence can match well with the published sentences of the target conference''. 
Based on the predictions of all sentences, we further use algorithms to integrate all sentences' predictions into the writing reviews.



Given this AI task, we deem that conversational XAI system should be prepared to \textbf{answer a wide range of knowledge gaps between the users and the AI models}~\citep{miller2017explainable}. 
That says -- the conversational XAI system is able to answer a variety of XAI questions that cover different perspectives of the system, including AI models, datasets, training and inference stages and even system limitations, etc~\citep{shen2021explaining}. 
Therefore, we design the XAI questions around four explanation goals, as illustrated in Table~\ref{fig:xai_questions} (1),
(a) \emph{understanding data}, which uses data to help contextualize users' understanding of where they abstract sit in the larger distribution; 
(b) \emph{understanding model}, which provides information on the underlying model structure so users can assess the model reliability;  
(c) \emph{understand instance}, which allows users to ask questions that dive into, each individual prediction unit (\ie sentence).
(d) \emph{improve instance}, which goes one step further than understanding, and targets the goal of helping people to \emph{improve} their writing by suggesting potential changes.
%
%
%
%


Embodied with the aforementioned two AI writing models and 8 types of AI explanations, we build up a preliminary system of conversational AI explanations for scientific writing support.  
The front-end user interface looks similar to Figure~\ref{fig:teaser}, which includes a \emph{human-AI task} panel on the left where users can inspect and edit their abstracts, and a \emph{conversational XAI} panel on the right where users interact with the XAI agent.
%
%
%
%
%
%
In the \textbf{human-AI writing task} panel, 
users can iteratively edit their abstracts, and submit them to receive AI assessments on their writing structure and style.\footnote{As the writing models of the preliminary and formal conversational XAI systems are identical, we encourage readers to refer to Section~\ref{subsec:writing_models} for more details of all the writing models and reviews.} 
%
%
%
As for the \textbf{conversational XAI} panel, at the initial entry, the panel provides a summary of the recommended edits (Figure~\ref{fig:pilot}A).
Then, as participants dive into each individual sentence, we allow them to select XAI methods they might find suitable by clicking on the corresponding buttons (Figure~\ref{fig:pilot}B).
The button-based design is inspired by the standard interface for service chatbots~\citep{yeh2022guide}, while participants were still allowed to just type their own questions.
This setting is also similar to the existing XAI interactive dialogue systems~\citep{tsai2021exploring,sun2022exploring}, where they provide different formats of AI explanation for the same prediction and evaluate human assessment on different explanations.

%
\subsubsection{Participants and Study Procedure.}
We recruited seven participants with diverse research backgrounds and experiences in the formative study: 1 assistant professor, 2 Ph.D. students, 3 industry scientists or engineers, and 1 master's student working on HCI, NLP, and AI research (refer to Table~\ref{fig:formative_table} for detailed demographic statistics).
%
%
%
The formative studies are conducted virtually via virtual conference calls on Zoom. 
During this study, participants were asked to either bring one of their abstract drafts or use one example provided by us. 
We conducted a semi-Wizard-of-Oz (WoZ) process where we encouraged users to think aloud during asking AI explanations to the XAI agent, with keeping in mind the goal of improving their abstract writing.
One researcher, who had several years of HCI and algorithmic AI explanation experience, acted as the XAI agent in this WoZ setting.
We collected users' reflections on the system and summarized them into design rationales below.

\subsection{Design Rationales}
\label{sec:principles}

While formative study participants all appreciated the access to multiple XAI methods, merely listing all XAI options for human use is not enough. Instead, they were frequently overwhelmed by the large number of options available. 
We combine their feedback with theoretic linguistic properties of human conversation~\citep{jurafsky2000speech,hutchby2008conversation}, and propose the following for design requirements for \convxai systems:

\begin{enumerate}[labelwidth=*,leftmargin=1.8em,align=left,label=R.\arabic*]
\item \label{req:multifacet}
    \textbf{Multifaceted}: \convxai system should provide diverse types and formats of AI explanations for users to choose from, and use multi-modal visualization techniques to display the explanations efficiently. As we have argued in Section~\ref{sec:formative}), to satisfy diver users needs~\citep{liao2020questioning,shen2021explaining}, it is imperative to \textbf{provide multiple XAI types and formats}. 
    Nevertheless, some formative study participants noticed that having all the explanations displayed at once is overwhelming, and preferred to have a ``overview first, details on demand'' structure~\citep{shneiderman2003eyes}.
    I-6 discussed that \emph{``I can tell the system knows a variety of AI explanations. However, it can be too much for me to understand all these explanations at once. I would prefer to know the ‘big picture’ first, and then drill down with ‘some options’ as I need to dive deeper.''}

\item \label{req:mix_init}
    \textbf{Mixed-initiative}: \convxai system should enable both user and XAI agent to initiate the conversation. Especially, it should proactively speculate the XAI user needs and prompt with next-step suggestions.
    One unique characteristic of conversations is mixed-initiative, \ie who drives the conversation~\citep{jurafsky2000speech}. Just as many existing conversational systems, we aim to mimic human-human conversations where initiative shifts back and forth between the human and the \convxai. This way, not only can the system answer users' questions, but it can also occasionally steer the conversation in different directions. In our study, we also found this to be quite essential, especially when users do not have a clear goal in mind (\eg \emph{``Which sentence in the abstract should I look into first?''}).
    
\item  \label{req:drilldown}
    \textbf{Context-aware drill-down}: \convxai system should allow users to drill down AI explanations with multi-turn conversations with awareness of the context.
    Linguistic theories model human conversation as a sequence of turns, and conversational analysis theory~\citep{hutchby2008conversation} describes the complex dialogues as joining the basic units, named adjacency pairs.
    This was also empirically validated in our pilot study. For instance, I-2 discussed potentially switching between explanations based on current observations: \emph{``I might directly ask the system how to rewrite the sentence to change this sentence into the background aspect ({\ie} ``counterfactual explanation''). But if its rewritten sentences are not good enough, I would check the most similar examples of background aspects to learn their style and write on my own then ({\ie} ``similar examples'')''}.
    Carrying over context throughout the conversation without users repeating themselves too much is useful for making the conversation natural and continuous.

\item \label{req:control}
    \textbf{Controllability}: \convxai system should be able to generate customized AI explanations that can satisfy the user needs and context.
    This includes both only displaying explanations that are relevant to their questions (\eg answer ``why this prediction'' with feature attribution), and adjusting the explanation settings (\eg number of important words to highlight).
    As I-7 said -- \emph{``I spent too much time on figuring out what each XAI means, then I forget what I want to write in the abstract. It would be great if to give me the AI explanations targeting my question and enable me to input some variables to generate the XAIs I want.}
    At the same time, users still preferred to have a default explanation first and then provide options to control the variables or diver deeper into details, so they only need to pay attention to parts that are worthy of personalization.
    \end{enumerate}


\section{\system
\hspace*{-0.06in}
\img{figures/logo-new-wotext.png}}
\label{sec:system}

\begin{figure*}
  \includegraphics[width=0.95\textwidth]{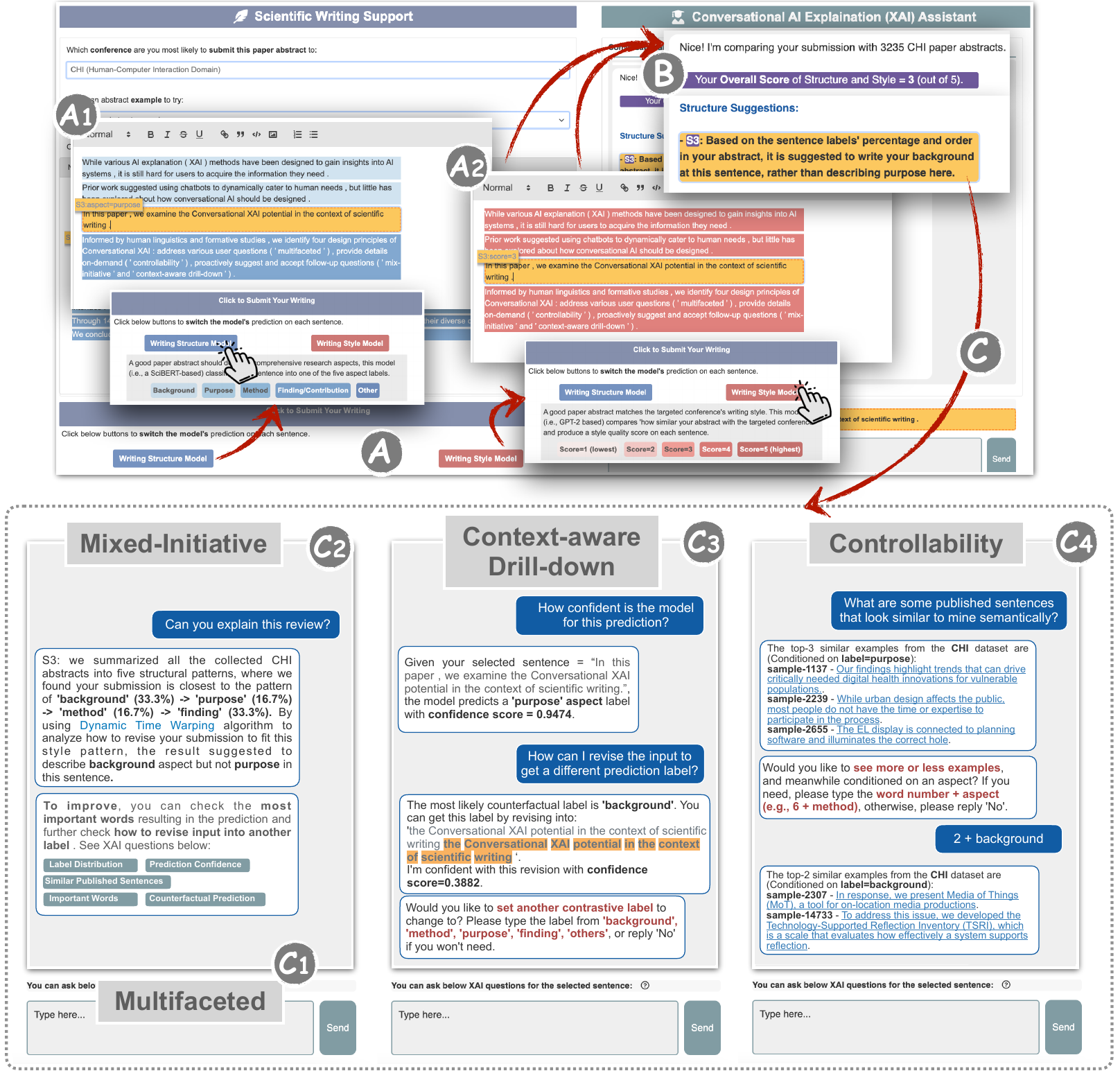}
  \caption{An overview of \system system. \system includes two writing models (A) to generate writing structure predictions (A1) and writing style (A2) predictions.
  Furthermore, the XAI agent in \system provides integrated writing review (B) followed by conversations with users to explain the writing predictions and reviews. Especially, the dialogue flows are designed to follow the four principles of \emph{``multifaceted''} (C$_1$), \emph{``mixed-initiative''}(C$_2$), \emph{``context-aware drill-down''}(C$_3$) and \emph{``controllability''}(C$_4$).
  }
  \label{fig:formal_convxai}
\end{figure*}


Based on the use scenario and design principles, we present \system, a system that applies conversational AI explanations on scientific writing support tasks, which incorporates the four rationales into the system design. The system aims to leverage conversational AI explanations on the AI writing models to improve human scientific writing. 
We extend the system developed in the formative study, which consists of a writing panel and an explanation panel. 
The writing panel is similar to the formative study, which can enable users to iteratively submit their paper abstract and check the writing model predictions for each sentence. We introduce more details of the scientific writing task and how the two writing models generate predictions and reviews in Section~\ref{subsec:writing_models}.
On the other hand, we significantly improve the conversational AI explanation panel by incorporating the four design rationales described above (Section~\ref{subsec:expl_panel}). Below, we elaborate on the ten formats of AI explanations included in our \system system, how we design the conversational XAI with the four principles, and the implementation of the system pipeline with details (Section~\ref{subsec:unified_xai}).

\subsection{Overview of User-Oriented \system Design}
\label{subsec:expl_panel}

%
The final \system user interface is illustrated in Figure~\ref{fig:formal_convxai}. 
We significantly revise the underlying dialog mechanism based on the preliminary system according to the four design rationales, so users can interact more smoothly with the XAI agent to cater to user demands.
We use Figure~\ref{fig:formal_convxai}C to demonstrate the design.

To design \system to be \textbf{mixed-initiative} (\ref{req:mix_init}), we start the explanation dialog with a review summary of the writing structure model and style model's outputs (Figure~\ref{fig:formal_convxai}B).
The users can select any one sentence (in this case, the third sentence with the sentence id \texttt{S3}) in this suggestion list to dive in, and start a conversation session on the sentence. 
Uniquely, to maintain \textbf{multifaceted} explanations (\ref{req:multifacet}) without overwhelming users, we add an additional explanation type, \emph{understand suggestion} --- answering questions like \emph{``Can you explain this review''} --- which provides general contextualization on a given suggestion (Figure~\ref{fig:formal_convxai}C$_2$).
To make it serve as proactive guidance towards more sophisticated XAI methods, the agent also initiates a prompt message \emph{``to improve...''} with a subset of relevant XAIs, based on the ``guess'' that users would want to improve their writing at this point.

To enable \textbf{context-aware drill down} (\ref{req:drilldown}), the user questions as well as agent answers are considered subsequently.
For example, in Figure~\ref{fig:formal_convxai}C$_3$, the user receives a review suggesting to describe \emph{background} aspect instead of \emph{purpose} aspect for the selected S3. 
The user firstly wants to know \emph{how confident the model makes this prediction.} 
Given the model confidence is quite high (around 0.95), she wanted to know how much she has to change in order to receive a different label.
The agent directly contextualizes these questions based on the suggested change in Figure~\ref{fig:formal_convxai}C$_2$ (``suggested to describe \textbf{background}''), and responds with a rewrite for the label \emph{background} without having to double-check with the user first.

Still, the default may not reflect users' judgment in some cases.
To mitigate potential wrong contextualization, we make the agent always proactively initiate hints for \textbf{controllablity} (\ref{req:control}), \eg ``would you like to...'' at the bottom of Figure~\ref{fig:formal_convxai}C$_3$.
Figure~\ref{fig:formal_convxai}C$_4$ provides a more concrete example: when the user asks for similar sentences published in the targeted conference, the XAI agent responds to the top-3 similar examples conditioned on the predicted aspect (\emph{i.e.,} \emph{purpose}) by default. 
However, as the user is suggested to rewrite this sentence into \emph{background}, she requests for the top-2 similar sentences which have \emph{background} labels by specifying ``2 + background'', so to use those examples as gold ground truths for improving her own writing.


\subsection{Human-AI Scientific Writing Task}
\label{subsec:writing_models}
%
We aim to provide two sets of writing support: (1) whether the abstract follows the typical semantic structure of the intended submission conferences, and (2) whether the abstract writing style matches with the conference norm.
To do so, we leverage two large language models to generate predictions for each abstract sentence. 

First, we use a \textbf{writing structure} model to assess the semantic structure by assessing if the abstract sufficiently covers all the required research aspects (\eg provide background context, describe the proposed method, etc.)~\citep{huang2020coda} (Figure~\ref{fig:formal_convxai}A$_1$).
We create the model by finetuning SciBERT-base~\citep{beltagy2019scibert}, a pre-trained model specifically captures scientific document contexts, on the CODA-19 datasets~\citep{huang2020coda}, which annotates each sentence in 10,000+ abstracts by their intended aspects, including Background, Purpose, Method, Finding/Contribution, and Other in the COVID-19 Open Research Dataset. The model achieves an F1 score of over 0.62 for each aspect and an overall accuracy of 0.7453.
The model performance is demonstrated in Appendix~\ref{appdx:model_performance}A.

While this model provides per-sentence predictions, the quality of an abstract depends more on the \emph{sequence} of sentence structures.
For example, ``background'' sentences should not be too many and should be primarily before ``purpose'' and ``method''. 
To support abstract improvement, we further implement a pattern explanation wrapper on top of the model, which suggests writers change some sentences' aspects to reach a better aspect pattern.
For example, ``background'' sentences should not be too many and should be primarily before ``purpose'' and ``method''. 
Therefore, we provide structure \emph{pattern} assessment, which suggests writers change some sentences' aspects to reach a better aspect pattern.
Specifically, for each conference (\emph{e.g.,} ACL), we clustered all abstracts in the conference into five groups and extracted the centers' structural patterns as the benchmark (\emph{e.g.,} ``background'' (33.3\%) -> ``purpose'' (16.7\%) -> ``method'' (16.7\%) -> ``finding'' (33.3\%)). Afterward, we compare the submitted abstract's structural pattern with the closest pattern using the Dynamic Time Warping~\citep{muller2007dynamic} algorithm to generate the structure suggestion for writers. See the extracted structural patterns for all conferences in Appendix~\ref{appdx:model_performance}B.

Second, we use a \textbf{writing style model} to predict the style quality score for each sentence, and check if the writing style matches well with the target conference.
As we intend first to support abstract improvement in the CS domain, we collect 9935 abstracts 
published during 2018-2022 from three conferences with relatively diverse writing styles, namely ACL (3221 abstracts), CHI (3235 abstracts), and ICLR (3479 abstracts), which are representatives of the top-tier conferences in Natural Language Processing, Human-Computer Interaction, and Machine Learning domains. 
More data statistics of the three conferences are in Appendix~\ref{appdx:model_performance}C.
To represent raw writing style match,  we use the style model to assign a perplexity score~\citep{jelinek1977perplexity} for each sentence, which is a measurement that approximates the sentence likelihood based on the training data.
Further, since the perplexity score is quite opaque, we add a normalization layer for better readability.
Specifically, we categorize the quality scores into five levels (\emph{i.e.,} score = 1 (lowest) to 5 (highest)), which is similar to the conference review categories that writers are familiar with. 
To achieve these five levels, for each conference, we got the distribution of all sentences' perplexity scores, and computed the [20-th, 40-th, 60-th, 80-th] percentiles of all the scores, then divided all scores based on these percentiles. See the quality score distribution in Appendix~\ref{appdx:model_performance}D.

To provide better overviews, we further offer an overall, abstract-level assessment by averaging its ``overall style score'' and ``overall structure score''. 
The ``overall style score'' is computed by averaging all sentences' quality scores. Whereas we compute the ``overall structure score'' as $\texttt{overall\ structure\ score} = 5 - 0.5 * \texttt{\#structure\ comments}$, where $\texttt{\#structure\ comments}$ means the number of structure reviews.


\subsection{A Unified Interface for Heterogeneous XAIs via Conversations}
\label{subsec:unified_xai}

\subsubsection{\system conversational XAI pipeline.}
\mbox{}\\
We develop the \system system to include a web server to host the User Interface (UI), and a deep learning server with GPUs to host both the  writing language models and AI explanation models. We mainly describe our implementation of the conversational XAI agent module below.
Specifically, we develop the conversational XAI pipeline from scratch based on the Dialogue-State Architecture~\citep{adi2016fine} from the task-oriented dialogue systems. The pipeline consists of four modules including a \emph{Natural Language Understanding} module that classifies each XAI user question into a pre-defined user intent, which is mapped into one type of XAI algorithm. The second module, named \emph{AI Explainers} is for generating ten types of AI explanations. Then the output is connected to the third module, named \emph{Natural Language Generation}, to generate natural language responses that are friendly to users. 
On top of the pipeline, we include a Global XAI State Tracker, to record users’ turn-based conversational interactions, including user intent transitions and the users' customization on AI explanations.
We introduce more implementation details below.

\begin{itemize}
    \item \textbf{Natural Language Understanding (NLU).} This module aims to parse the XAI user question and classify the user intent into which types of AI explanations they may need. 
    We currently design the intent classifier to be a combined model of a rule-based classifier and a Deberta-based model. 
    We trained the Deberta-based classifier~\citep{he2020deberta} to do the intent classification, where we classify each user question into one of the eleven pre-defined XAI user intents (\ie ten user intents and the ``others'' type).
    
    \item \textbf{AI Explaners (XAIers).} Based on the triggered XAI user intent, this module selects the corresponding AI explainer algorithm to generate the AI explanations. Currently, we implemented the \textbf{AI Explainers} to include ten XAI methods to answer the ten XAI user questions listed in Table~\ref{fig:xai_questions} correspondingly.
    Furthermore, we design a unified API to generate heterogeneous AI explanations to implement this \emph{AI Explainer},
    which can incorporate the four principles discussed above. For example, the \emph{AI Explainers} enables users to input the personalized variable (\eg how many similar examples to explain) they need, and the \emph{AI Explainers} will feed the ``user-defined'' variable into the AI algorithm to generate ``user-customized'' AI explanations.

    \item \textbf{Natural Language Generation (NLG).} Given the outputs from the \emph{AI Explainers}, we leverage a template-based NLG module to convert the generated AI explanations into natural language responses. Note that we especially design the NLG templates to be multi-modal, so that it enables both free-text responses and visual-assisted responses (\eg heatmap to explain feature attributions) to meet users' needs.

    \item \textbf{Conversational XAI State Tracker.} As our \system empowers users to choose from multiple types of XAI methods, drill down to AI explanations and make XAI customizations. We specifically design the global Conversational XAI State Tracker to record users’ turn-based conversational interactions. Particularly, we record the turn-based user intent transitions and the users' customization on AI explanations.
\end{itemize}

Overall, we design the conversational XAI pipeline to be model agnostic and XAI algorithm agnostic. This enables the \system system to be naturally generalized to various AI task models and AI explanation methods. 

\subsubsection{Embodying Heterogenous AI Explanations in \system.}
\mbox{}\\
Here, we provide technical details on all the explanation methods enumerated in Table~\ref{fig:xai_questions}.
First, \textbf{understanding data and model} requires more global explanations that summarize the training data distribution as well as the model context.
For the data, we include data sheets~\citep{gebru2021datasheets} for the datasets used. We further compute important attribution distributions, including the quality scale mentioned above, the structure label distribution, and the sentence length. Such information also helps users contextualize where their abstract sits on the distribution.
Similarly, for providing sufficient model information, we incorporate model cards~\citep{mitchell2019model} for SciBERT and GPT-2, and adjust them based on our finetuning data.

Second, for understanding and improving models, we leverage the state-of-the-art XAI algorithms to generate local AI explanations. This includes: 
\begin{itemize}
    \item \textbf{Prediction confidence}, which is the probability score after the softmax layer of the SciBERT model reflecting model prediction certainty. This explanation is only provided for the writing structure model. 
    
    \item \textbf{Similar examples}, which retrieves semantically similar sentences published in the target conference to be referenced. We assess this with the dot product similarity of the sentence embeddings~\citep{pezeshkpour2021empirical} (derived from the corresponding writing assistant models). This is provided for both writing structure and style models. \footnote{Note that we deem similar examples useful mostly because users also tend to learn about the writing academic writing styles through mimicking published papers, but whether such reference counts as (or encourages) plagiarism is an open question that needs investigation.}
    
    \item \textbf{Important words}, which aims to highlight the top-K words that attribute the writing model to the sentence prediction. We leverage the \emph{Integrated Gradient approach}~\citep{mudrakarta2018did} to generate the word importance score (\emph{i.e.,} attribution).

    \item 
    \textbf{Counterfactual Predictions}, which re-writes the input sentence with a desired aspect while keeping the same meaning.
    We design an in-context learning approach using GPT3~\citep{gpt3} to re-write sentences. Given an input sentence, we first retrieve the top-5 semantically similar sentences for each of the five aspects from the collected CS-domain abstracts (the semantic similarity between sentences is measured by the cosine similarity over sentence embeddings~\citep{reimers-gurevych-2019-sentence}). A total of 25 examples would be extracted dynamically and form a prompt using the template ``\{example sentence\} is labeled \{aspect\}''. After showing 25 examples, we add ``Rewrite \{input sentence\} into label \{desired aspect\}'' to the prompt. GPT3 then follows the instruction to generate a modified sentence with the desired aspect label.
    
\end{itemize}

Finally,  as described in Section~\ref{subsec:expl_panel}, we further add \textbf{understanding suggestions} to answer the general question of ``\emph{how did the system generate the suggestions?}'', and provide pointers to other finer-grained explanations methods. 
We create ``suggestion explanations'' for each piece of writing feedback. Particularly, we create one template for writing structure review, writing style review, and sentence length review, respectively. 
In each template, we describe how we compare all predictions in the abstract with the target conference data statistics to generate the corresponding review. Then we initiate an ``improving message'' aiming to guide users in how to use XAI to improve their writing, this message includes the buttons of potential XAI methods that we deem users might use for resolving this review (as one example shown in Figure~\ref{fig:formal_convxai}).

\subsection{Implementation Details}
\label{subsec:implementation}

We develop \system as a stand-alone system independent of any platforms. The front-end of \system is built on the open-source Flask codebase with HTML, CSS, and Javascript codes hosted on a web server. On the other hand, the back-end of \system is a deep learning server with GeForce RTX 2080 GPUs hosting AI writing models and the conversational pipeline to generate heterogeneous AI explanations in Python and PyTorch. We also refer to ParlAI~\citep{miller2017parlai} to develop the conversational AI pipeline in \system. The front-end and back-end of \system communicate with the WebSocket protocol using the Socket.IO library and save all \system data in the MongoDB database. Around 4,300 lines of font-end codes and 6,500 lines of back-end codes are added, resulting in around \textbf{10,800 lines} of code in the final \system.
Furthermore, to better generalize the unified API for conversational XAI for future study, we \textbf{extract the core unified API in \system into a Notebook}\footnote{\label{note-api}See the unified API of conversational XAI at: \url{https://github.com/huashen218/convxai/blob/main/notebook_unified_XAI_API/convxai_unified_api.ipynb}} for further research reference.

\begin{figure*}
  \includegraphics[width=0.95\textwidth]{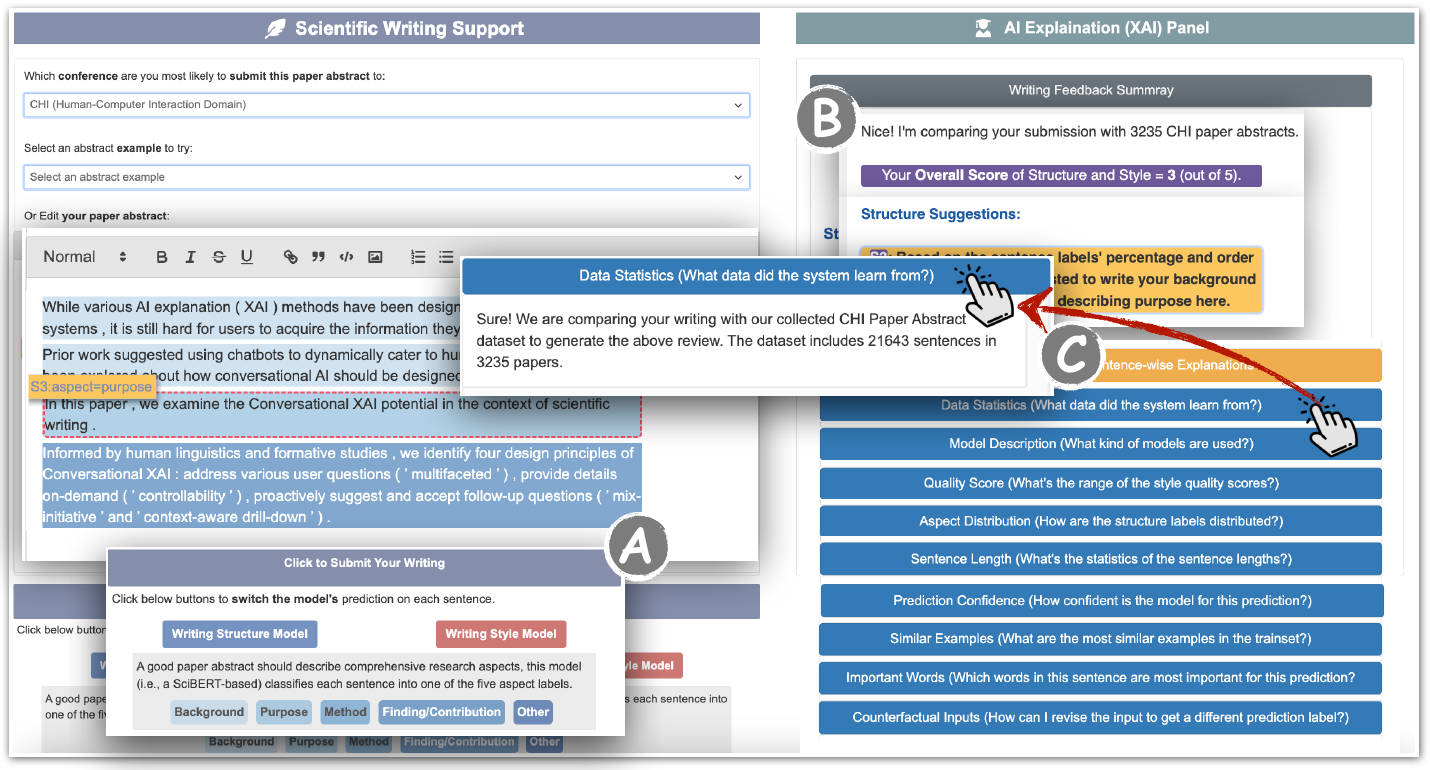}
  \caption{An overview of \baseline system. Similarly, it includes (A) two writing models to generate writing structure predictions, and (B) integrated writing review followed by (C) static XAI buttons to show and hide the explanations.
  }
  \label{fig:baseline}
\end{figure*}

\section{User Studies}
\label{sec:eval}

We conducted two within-subjects human evaluation studies, where we compare the proposed \system against \baseline, a GUI-based universal XAI system. 
The user study aimed to investigate how users leverage the XAIs systems to better understand the AI writing feedback and improve their scientific writing. We particularly designed the study to consist of (1) an open-ended writing task to evaluate the effectiveness of user-oriented design in the system, and (2) a well-defined writing task to investigate how systems can help users improve their scientific writing process and output in practice. 
Specifically, we pose the following research questions:
\begin{itemize}
    \item \textbf{RQ1}: Can user-oriented design in \system help humans better understand the AI feedback and perceive improvement in writing performance?
    \item \textbf{RQ2}: Can the \system be useful for humans to achieve a better writing process and output?
    \item \textbf{RQ3}: How do humans leverage different AI explanations in \system to finish their practical tasks?
\end{itemize}

\subsection{Task1: Open-Ended Tasks for System Evaluation}
\label{sec:task1}

\emph{Can \system help users to better understand the writing feedback and improve their scientific writing?}
\emph{What designs support this purpose?}
With these questions kept in mind, we conduct a within-subject user study comparing \system with a \baseline baseline interface.  
Following the study, we ask participants to comment on the systems and examine how they use the \system to improve their writing by observing their interaction process.

\subsubsection{Study Design and Procedure}

\paragraph{\textbf{\emph{Participants and \baseline System.}}}

We recruited 13 participants from university mailing lists. All the participants had research writing experience, resided in the U.S. and were fluent in English. 
The group has no overlap with the formative study participants, none of them had used \system prior to the study.
Each study lasted for one and a half hours. The participant was compensated with \$40 in cash for their participation time.


We ask each participant to compare \system with a baseline system, named \baseline, shown in Figure~\ref{fig:baseline}. 
The \baseline system also consists of all the AI explanation formats included in \system. However, it statically displays all the XAI formats on the right-hand view panel instead of using dynamic conversations to convey XAIs.
To display all the XAI for each sentence, users can select a sentence from the left writing editor panel to be explained, then generate all XAI formats by clicking a trigger button at the right panel. As a result, users can view all XAI formats with each having a button to control hiding and showing the AI explanations results. In other words, \baseline remains multifaceted (\ref{req:multifacet}) and somewhat controllable (\ref{req:control}), but does not have drill-down (\ref{req:drilldown}) or mixed-initiative properties (\ref{req:mix_init}).

\paragraph{\textbf{\emph{Study Procedure.}}}

We conducted \emph{within-subjects study} where we have the same users to interact with both the proposed \convxai system and \textsc{SelectXAI} baseline system. 
Each user study consists of three steps where \emph{i)} we first instruct each user \emph{how to use the \system and \baseline systems} by showing them a live demo or recorded videos. They can stop the instruction anytime and ask any questions about the tutorials. \emph{ii)} After the system tutorials, we invited the users to explore both \system and \baseline systems with the pre-defined order. Particularly, we randomized the orders of all 13 studies. As a result, we ask 7 participants to start with the \system group, and 6 participants to start with the \baseline group.
\emph{iii)} Finally, we ask the users to fill in a post-hoc survey including two demographic questions and 14 questions rating their user experience on 5 points Likert scale. We further ask them three open-form questions after the survey to interview their opinions about the \system and \baseline systems.

During the step \emph{ii)} and \emph{iii)}, we recorded the video of the process, and encouraged them to think aloud.
Besides, we designed the users to evaluate two systems either both with their own papers or both with the examples we provide. We encouraged users to use their own paper drafts where users had more incentives to improve their writing. As a consequence, 12 out of 13 users submit their own drafts or published papers.

\subsubsection{Study Results}
\label{sec:result}

\begin{figure*}
    \includegraphics[trim={0 23cm 25cm 0cm}, clip, width=1\textwidth]{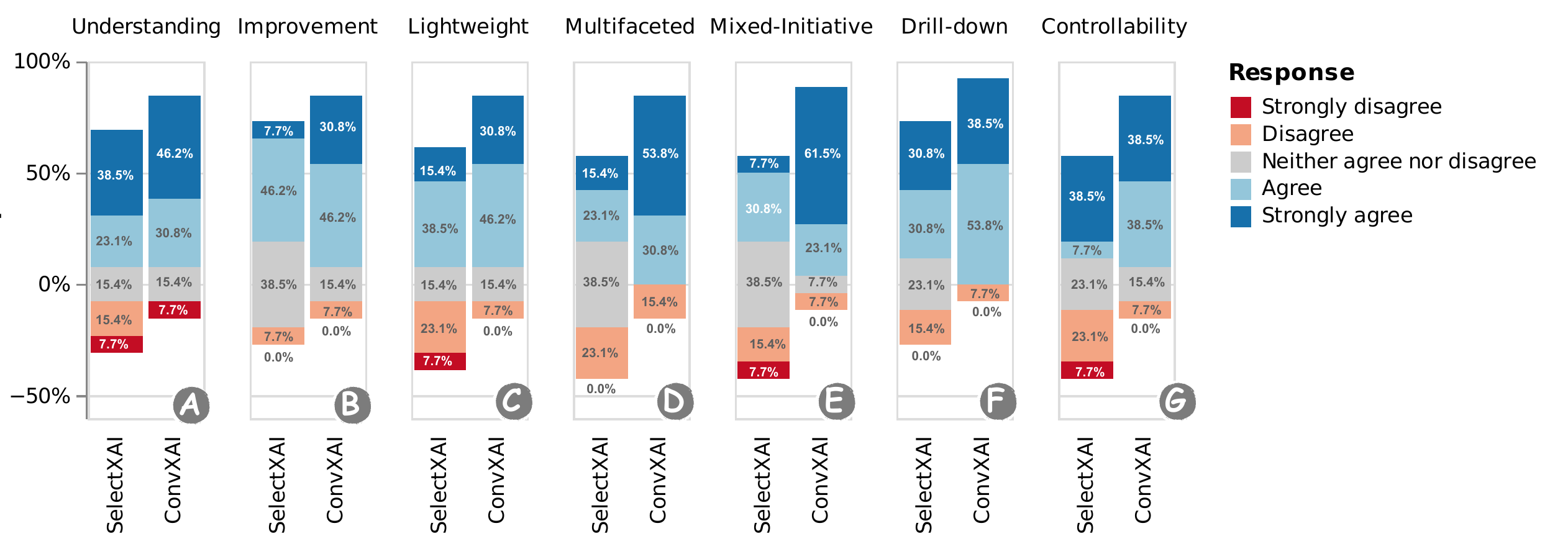}
  \caption{Analyses on users' self-ratings on their experiences playing with \system and \baseline. They self-rated \system to be better on all dimensions, and most significantly on the usefulness of mix-initiative and multifaceted functionality.}
  \label{fig:user_study}
\end{figure*}

\mbox{}

We first look into the overall usefulness of \system, and answer the question: is \system useful for users' ultimate goal of understanding and improving their abstract quality (\emph{RQ1})?
We summarize participants' ratings on the two systems, \system and \baseline, in Figure~\ref{fig:user_study}.
We performed the non-parametric Wilcoxon signed-rank
test to compare users' nominal Likert Scale ratings and found that participants self-perceived \system to \textbf{help them to better understand why their writings were given the corresponding reviews} 
(\system $4.07 \pm 1.18$ vs. \baseline $3.69 \pm 1.37$, $p=0.036$, Figure~\ref{fig:user_study}A).
.
They also felt that \textbf{\system helped them more in improving their writing} ($4 \pm 0.91$ vs. $3.53 \pm 0.77$, $p=0.019$, Figure~\ref{fig:user_study}B).
The helpfulness are likely because participants can more effectively find answers to their diverse questions, which we detail in Section~\ref{subsec:strength}.

Besides their promising self-reflection, 3 out of 13 participants actually edited and iterated their abstracts in \system. 
They all successfully addressed the AI-raised issue (\ie the corresponding suggestion disappeared when they re-evaluated the edited version).
However, the other 10 participants showed low incentive to revise the published abstracts.
Through interviews, we summarize some challenges they faced in interacting with the current \system in Section~\ref{sec:limitation}.
\label{subsec:strength}
Through the study observations and free-form question interviews with users, we obtained that 9 out of 13 participants prefer to use \system than \baseline system for improving their scientific writing. 
We conjecture that this might primarily result from \system's ability to answer user questions more \emph{sufficiently}, \emph{efficiently}, and \emph{diversely}.
More specifically, the benefit comes from three dimensions:

First, \textbf{\system reduces users' cognitive load digesting the available information}.
9 participants were overwhelmed by \baseline, and complained that they had to manually click through all the available buttons before they realize all of them contain explanations in the exact same sentence.
In contrast, \system releases the \emph{same} information more \emph{gradually} through the back-and-forth conversations.
Participants especially appreciated that the initial suggestions from \system (mixed-initiative, \textcolor{principle}{\textbf{\texttt{R2}}}
), as it enables them to interact with the system without having to understand its full XAI capability (unlike in \baseline).
For example, P12 pointed out, \emph{``it is very helpful that the XAI agent can give me some hints on using the AI explanations. Especially when I'm a novice of scientific writing and AI explanation knowledge, this helps me get involved in the system more quickly.''}
Indeed, this is also reflected in participants' ratings: in Figure~\ref{fig:user_study}E, participants found \system helped them figure out how to inquiry about a sentence (\system$ 4.23 \pm 0.83$ vs. \baseline $3.77 \pm 1.09$, $p=0.001$).
Additionally, it is important that the \system is robust in detecting user intents, such as being tolerant of user input typos. As P1 and P2 mentioned, ``I really like the \system that allows my typos by only capturing the keywords, so that I don't need to memorize much knowledge for using the system.''

Second, \textbf{\system enables users to pinpoint the XAI questions efficiently.}
We quantified the types of questions participants frequently asked, and found 9 out of 13 participants had explicit preferences for using some specific AI explanations formats. Among these 9 users, 66.67\%, 55.56\%, and 33.33\% participants primarily used \emph{counterfactual explanation}, \emph{similar example}, and \emph{feature attribution} explanations, respectively.
This suggests that, indeed, people have different kinds of questions and XAI needs.
Participants liked that they could take the initiation and prioritize their own needs, and simply query the associated XAI through the dialog, 
whereas in \baseline, ``I just go over all the explanations and read everything, for some of the explanations I just don't care, this is somehow a bit overwhelming to me.'' (P3)
This also means they were much less likely to be distracted by duplicate details (\eg
P1: \emph{``I only need to understand the general information about the model and data at the very beginning, after that, I don't need to check it repeatedly every time for each sentence.''}
), or explanations irrelevant to their questions.
As a result, they rated \system to provide explanation more easily and more naturally (
\system$ 4.0 \pm 0.91$ vs. \baseline $3.3 \pm 1.25$, $p=0.008$, Figure~\ref{fig:user_study}C).

Interestingly, having users to self-initiate questions brought an unexpected benefit --- it helps users think through the writing and what they actually want to understand. As P6 said, \emph{``Compared with \baseline, \system slows down the interaction and gives me the time and incentive to think about what I want the robot to explain.''} P4 also pointed out, \emph{``The follow-up hints inspire me to think more about how to use the XAI for my writing.''}
This somewhat echoes prior work that showed pairing humans with slower AIs (that wait
or take more time to make recommendations) may provide humans with a better chance to reflect on their own decisions~\citep{park2019slow}.

Third, \textbf{\system provides sufficient AI explanations crafted for user need}. 
Interestingly, though \system and \baseline implemented the same amount of explanation types and participants were overwhelmed by \baseline, they still rated \system to have a more sufficient amount of explanations (multi-faceted, \system $4.23 \pm 1.09$ vs. \baseline $3.31 \pm 1.03$, $p=0.007$, Figure~\ref{fig:user_study}D).
\system's controllability (\system $4.08 \pm 0.95$ vs. \baseline $3.46 \pm 1.45$, $p=0.014$, Figure~\ref{fig:user_study}G) played an important role here (\system $4.07 \pm 0.95$ vs. \baseline $3.46 \pm 1.45$, $p=0.001$, Figure~\ref{fig:user_study}E).
Participants mentioned that it is essential for them to customize \emph{how} their questions were answered, and were satisfied that they could customize the level of details in one XAI type (\eg number of similar words in feature attribution, targeted label in counterfactual prediction, etc.), whereas \baseline did not provide the same level of control (as per \emph{status-quo}).
We observe all (13 out of 13) participants performed the personalized control on generating AI explanations during the user study.

The ability to drill down was equally important. 
We saw users performing different kinds of follow-ups based on their current explorations.
For instance, as P5 mentioned, \emph{``I would first check the \emph{model confidence} explanation, if the confidence score is low, I would directly ignore this sentence prediction which makes my writing much easier. However, if the confidence score is high, I will use the \emph{counterfactual explanation} to check how to revise this sentence.''}
Participants also mentioned ``the function of enabling users to generate these personalized explanations are the most important features'' resulting in why they prefer \system over \baseline systems. Like P8 pointed out, \emph{``I think \baseline has the advantage of easier to use because the learning curve is short. However, I would still prefer \system because it can provide me with much more explanations that I need.''}
To better understand users' preferences on explanations, we summarize some use patterns in \system in the next section.

\begin{figure*}
    \includegraphics[width=1\textwidth]{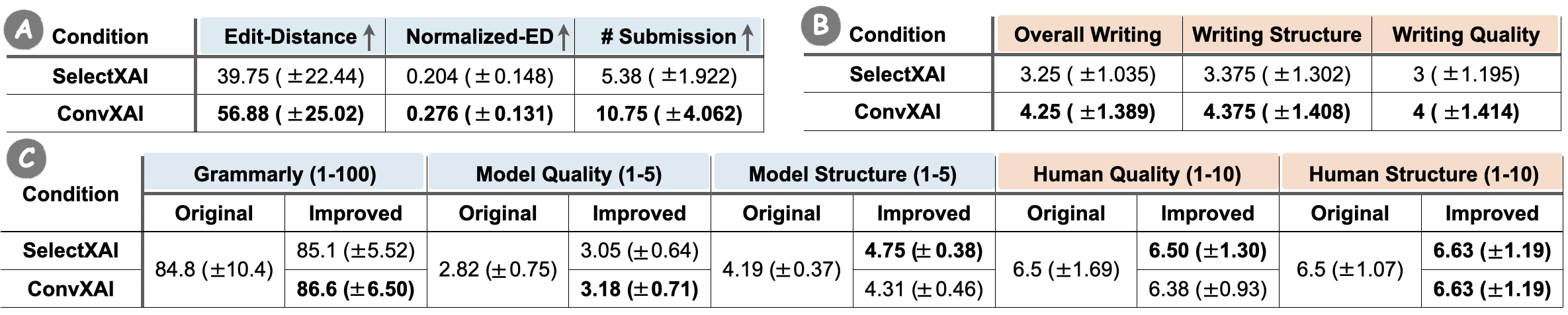}
  \caption{Evaluation of \textbf{Productivity} (A), \textbf{Perceived Usefulness} (B), and \textbf{Writing Performance} (C) measurements to assess users' writing performance in Task2. 
  (A) We deploy \textbf{Productivity} with three auto-metrics including ``Edit Distance'', ``Normalized-Edit-Distance'', and ``Submission Count''. 
  (B) We ask users to rate their perceived system usefulness for improving ``Overall Writing'', ``Writing Structure'', and ``Writing Quality''. 
  (C) We evaluate writing outputs using both auto-metrics (\ie ``Grammarly'', ``Model Quality'', and ``Model Structure''), and human evaluation (\ie ``Human Quality'' and ``Human Structure''). 
  }
  \label{tab:writing_eval}
\end{figure*}

\subsection{Task2: Well-defined Tasks for Writing Evaluation}
\label{sec:task2}

To answer RQ2, we further evaluate participants' productivity and writing output quality to assess the usefulness of \system and \baseline on human writing performance in Task 2. 
\subsubsection{Study Design and Procedure}
\paragraph{\textbf{\emph{Participants and Grouping.}}}
We recalled 8 users, who have joined Task1 and been familiar with the system, to participate in Task2 again. There are two reasons to recruit the same group of users again: i) the experience in Task 1 could help users reduce their learning curve and cognitive load on familiarizing the XAIs and systems. Therefore, users can focus more on the writing process;
ii) this design can potentially provide a temporal change in user behaviors on leveraging the systems. 
To conduct rigorous human studies, we divide 8 users into 4 pair of groups, with groups' research domains lying in ``NLP'', ``HCI'', ``AI'', and ``AI'', respectively.

\paragraph{\textbf{\emph{Study design and paper selection.}}}
Similar to Task1, we also conducted a within-subjects study, but with the objective of evaluating users' scientific writing outputs with the help of \system and \baseline systems.
For each group of two users, we ask them to rewrite the same two papers asynchronously, with a reverse order of system assistants. For instance, within the same group, user1 rewrites with `paper1-\system' followed by `paper2-\baseline' settings, whereas user2 rewrites with `paper1-\baseline' and `paper2-\system' settings successively. Hence, these settings eliminate the correlations between papers and system types and orders.
Afterward, we evaluate the users' writing outputs and experience with a set of metrics, including a real-human editor evaluation, a set of auto-metrics, and a post-survey.

For a fair comparison, we pre-selected eight papers (\ie 2 papers * 4 domain groups) for users to rewrite, which are recently submitted to arXiv (\ie around Nov/29/2022) within the domains of Artificial Intelligence\footnote{https://arxiv.org/list/cs.AI/recent.}, Computation and Language\footnote{https://arxiv.org/list/cs.CL/recent}, and Human-Computer Interaction\footnote{https://arxiv.org/list/cs.HC/recent}.
Also, we followed a set of rules during paper selection:
i) The papers are not in the top-5 best papers ranked by the editor and accepted by journals or conferences;
ii) Users don’t need specialized domain knowledge to improve writing. (e.g., no need to read the whole paper's contents to improve the writing);
iii) The AI aspect labels and quality score predictions are correct (checked by the authors).
During the study, we also recorded a video of the process and encouraged the participants to think aloud.

\subsubsection{Study Results.}
\mbox{}

We evaluate participants' scientific writing performance quantitatively in terms of \emph{productivity} and \emph{writing performance} (\ie how many changes have been made and whether the improved writing outputs are scored better). Akin to Task1, we also qualitatively assess participants' \emph{perceived usefulness} with 5 points likert scale from the post-survey.

\paragraph{\textbf{\emph{Productivity.}}} We evaluate \emph{productivity} with respect to the ``Edit-Distance'' and the ``Normalized-Edit-Distance'' (``Normalized-ED'') between the original paper abstract and the modified version from participants. We leverage Damerau–Levenshtein edit distance~\citep{levenshtein1966binary,damerau1964technique} and its normalized version~\citep{yujian2007normalized} to compute these two metrics. 
From Table~\ref{tab:writing_eval} (A), we observe that participants' edit distance using the \system is 43.09\% (\ie M=56.88 vs. M=39.75) higher than that using \baseline in average, meanwhile, the normalized edit distance is 35.29\% (M=0.276 vs. M=0.204) higher comparing \system and \baseline as well. This demonstrates that the \system is potentially useful to help users make more modifications to writing than that using the \baseline system.

Besides, we also record the ``Submission'' counts representing how many time the users modified their draft and re-submitted to the systems. Table~\ref{tab:writing_eval} (A) shows participants submitted 99.81\% more times with \system than using \baseline during the writing, with a statistically significant difference (p=0.0045). This result also indicates users tend to interact and submit more with \system than \baseline for rewriting the abstracts.

These findings are consistent with the users' think-aloud notes, in which most of them preferred to use the \system than \baseline for improving writing. Like P5 (who uses \baseline first followed by \system) mentioned, ``I somehow struggled with using the \baseline system because it provides very limited help. But I kind of started enjoying the writing process with the help of \system. ''

\begin{figure*}
\includegraphics[width=1\textwidth]{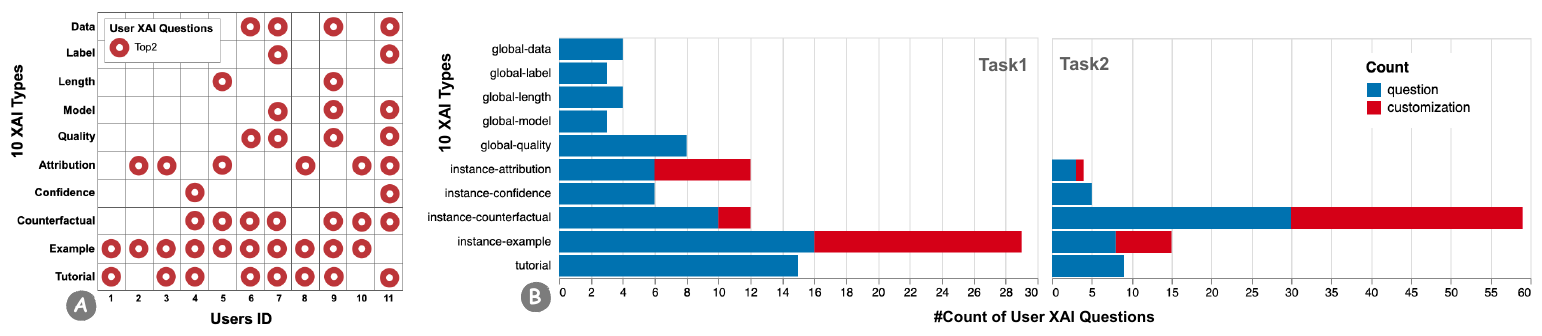}
  \caption{User demands analysis during using \system to improve scientific writing in Task 1 and Task 2. Particularly,
  (1) We ranked the top-2 most frequently requested XAI methods by each user ID in Task 1(A). 
  (2) We compute all the users' question amounts for each of the 10 XAI methods in (B) Task 1 and Task 2.}
  \label{fig:user_needs}
\end{figure*}

\paragraph{\textbf{\emph{Writing Performance.}}}
To understand whether \system can actually help users improve writing outputs, we compare the abstracts before (\ie Original) and after (\ie Improved) editing with \system and \baseline as shown in Table~\ref{tab:writing_eval}. We evaluated abstracts using three different measurements: 
(\textit{i}) Grammarly,
(\textit{ii}) \system's built-in models,
and (\textit{iii}) human evaluation.
To measure the abstract quality with Grammarly, we set Grammarly's suggestion goal as audience = expert and formality = formal, manually copy-and-paste all the abstracts to Grammarly, and record the scores.
Besides, we also adopt the two \system's built-in models, including the writing style model and the writing structure model. We leverage them to measure abstracts' language quality and abstract structure, respectively. These scores are also the AI scoring feedback for users during their writing tasks.
For human evaluation, we hire one professional editor to rate abstracts' quality in terms of language quality and abstract structure. Note that it is difficult to find an expert who is experienced in reviewing abstracts of all ``NLP'', ``HCI'', and ``AI'' domains. Therefore, we are also aware of the limitation of these human evaluations.

All scores are demonstrated in Table~\ref{tab:writing_eval} (C). We can observe that, by comparing with \emph{Original} scores, \textbf{both \system and \baseline are useful for humans to improve their auto-metric writing performance}, including the ``Grammarly'', ``Model Quality'', and ``Model Structure'' scores. Furthermore, \system specifically outperforms \baseline on Grammarly and writing quality metrics, indicating that \textbf{\system can potentially help users to write better grammar-based and style-based sentences} in scientific abstracts than \baseline.
On the other hand, the human editor's evaluation shows inconsistent results, where \textbf{\system and \baseline can both improve the writing Structure} evaluations, but not in the Quality metric. 
 %
To probe the inconsistency between human and auto-metric evaluations, we further compute the Pearson correlation between the model scores and the human ratings and find that both quality and structure are negatively correlated or not correlated (quality: -0.0311 and structure: -0.1150), 
showing that there is a misalignment between humans and models.

Therefore, we posit that both universal XAI systems, including \system and \baseline, are useful to improve human writing performance under auto-metric evaluations. Particularly, \system can outperform \baseline in terms of grammar and style-based writing quality.
Besides, as the human is not aligned with model evaluations based on Pearson correlations, the improvement failed in the human quality metric. This negative finding actually provides valuable insights into the importance of aligning the human judgment and model objective in AI tasks, so that users can use the systems to effectively reach both improvement goals.

\paragraph{\textbf{\emph{Perceived Usefulness.}}}
In the post-survey, we also ask users to rate their perception of system usefulness in terms of assisting their abstract writing. We particularly measured the users' perceived usefulness on ``Overall Writing'', ``Writing Structure'' improvement, and ``Writing Quality'' improvement. We design these three metrics to be consistent with the feedback from the AI writing models. Shown in Table~\ref{tab:writing_eval} (B), we can see participants perceived \system to be 1 (out of 5) point higher than \baseline in terms of use on all writing aspects. 
At the end of the survey, we further ask which AI explanations or system functions they perceived to be most useful, we elaborate on this finding in Sec~\ref{sec:use-pattern} below.

\subsection{Usage Patterns with \system}
\label{sec:use-pattern}

We propose \system based on the statement that universal XAI interfaces are important for satisfying user demands in real-world practice. 
In this section, we provide practical evidence to support that the \emph{universal XAI interface is indeed a necessary design of useful XAI for real-world user needs}.

By reviewing all 11 (from Task 1) and 8 (from Task 2) recorded study videos, we collected all the users' XAI question requests when they leverage \system to improve writing. In total, there are \textbf{95} and \textbf{92} XAI user requests in Task 1 and Task 2, respectively. 
Based on analyzing these XAI user requests., we demonstrate Figure~\ref{fig:user_needs} to provide detailed insights on practical user demands.
More specifically, in Figure~\ref{fig:user_needs} (1), we visualize each individual user's top-2 priority in using the different XAI methods. In Figure~\ref{fig:user_needs} (2), we accumulate all users' requests on each XAI method to visualize the usage distribution among the ten XAI methods. We also separately visualize Task 1 and Task 2 in order to observe the temporal usage patterns on XAI methods. We summarize our findings in detail below.

\subsubsection{\textbf{Different users prioritize different AI explanations and orders for their needs.}}

First, focusing on the same task but with different users, we observe that \emph{different users often prioritize different types of AI explanations even within the same task}.
In specific, for Task 1 shown in Figure~\ref{fig:user_needs} (A1),
although 9 users (\ie 1,2,3,4,6,7,8,9) prioritize using ``Examples'' explanations, the other 2 users (\ie 5,11) leverage ``Attributition'' and ``Confidence'' explanations most in their writing task 1. Besides, the 2nd-popular AI explanations of the 11 users are scattered among all the 10 XAI types without a unified pattern.

Additionally, in Task 2 with Figure~\ref{fig:user_needs} (B1), we can see users' top2 explanations are converging into instance-wise explanations (\ie ``Attribution'', ``Counterfactual'', etc). In specific, 7 out of 8 users prioritize ``Counterfactual'' and the other one leveraged ``Example'' explanation the most.
This is also consistent with the user's think-aloud observation. For instance, P5 lacks an AI background and didn't understand what ``Prediction Confidence'' means in this situation, whereas P11 mentioned \emph{``model confidence is the first explanation I'll ask to decide whether I'll ignore the prediction or continue the explanations.''}

Furthermore, we accumulate the users' XAI request counts for each XAI type and show the results of Task 1 and Task 2 in Figure~\ref{fig:user_needs} (A2) and (B2), respectively. We can observe that although user needs are often dominated by one XAI type (``Example'' and ``Counterfactual'' in Task 1 and 2, respectively), users also leverage \system to probe a wide range of other XAI types, such as ``XAI tutorial'', ``Confidence'', ``Attribution'', etc.) 
In short, these findings validate that it is important to use the universal XAI interface like \system, which can \textbf{accommodate different users' backgrounds and practical demands}.

\subsubsection{
    \textbf{User demands are changing over time.}}
In addition, we focus on the changes of user demands over time. We specifically compare the same user group's XAI needs in the two Tasks. 
By comparing Figure~\ref{fig:user_needs} (A1) vs. (B1), we can see that the top of users' XAI demands is gradually converging into the instance-wise explanations, including ``Counterfactual'', ``Example'', ``Confidence'', ``Tutorial'' and ``Attribution'' explanations. 

This can be further verified by comparing Figure~\ref{fig:user_needs} (A2) vs. (B2). We can see that i) user demands in Task 2 are highly skewed to ``Counterfactual'' explanations, which are two times more than the ``Example'' explanation ranked as top in Task 1.
ii) Users leverage much less and even no global information explanations (\eg ``Data'', ``Model'', ``Length'', etc) in Task 2. 
This is also consistent with the user think-aloud notes, where P4 pointed out ``After I know these data and model information, I might not need them again a lot, unless I need this information to analyze each sentence's prediction later.''

This again shows that it is important to design XAI systems to be a universal yet flexible XAI interface, as \system, to \textbf{capture the dynamic changes of user needs over time}.

\subsubsection{\textbf{Proactive XAI tutorials are imperative to improve the XAI usefulness.}}
\label{sec:xai_tutorial}
Both our pilot study and the two tasks illustrate that \textbf{providing users with instructions on how to use XAI is crucial}.
Particularly, 
echoing the ``Mixed-initiative'' design principle, we proactively give hints of XAI use patterns (\ie how to use AI explanations) for improving writing during the conversations. In Table~\ref{tab:use_pattern}, we exemplify a set of user patterns to resolve different AI writing feedback. 

From Figure~\ref{fig:user_needs} (A1) and (B1), we can observe that 72.73\% (8 out of 11) users and 37.5\% (3 out of 8) users prioritize ``Tutorial'' explanations as top-2 during Task 1 and 2, respectively.
Similarly, in Figure~\ref{fig:user_needs} (A2) and (B2), the accumulated counts of ``Tutorial'' explanations also ranked within top-3 in both Task 1 and 2, indicating a high user demand for checking tutorial/hints of XAI usage patterns.

Furthermore, we also observe a decreasing trend of ``Tutorial'' explanation needs over time by comparing Task 1 and Task 2. This potentially indicates that users are gradually being more proficient in using AI explanations for their own needs.

\subsubsection{\textbf{XAI Customization is crucial.}}
\label{sec:xai_customization}
By observing the think-aloud interviews in the two tasks, we deem one fundamental reason that \system outperforms \baseline is that it provides much more flexible customization for the user request. This corresponds to the ``Controllability'' design principle derived from the pilot study as well.
Note that we only design 3 out of 10 AI explanations to enable XAI customization. Particularly, we allow users to specify one variable (\ie ``target-label'') for generating ``Counterfactual'' and ``Attribution'' explanations, and four variables (\ie ``target-label'', ``example-count'', ``rank-method'', ``keyword'') to generate ``Example'' explanations.

Importantly, by visualizing Figure~\ref{fig:user_needs} (A2) and (B2), we observe that there are 22.11\% and 40.22\% practical user requests for XAI customization in Task 1 and 2, respectively.
Besides, all users in both Task 1 and 2 requested XAI customization during their studies. These findings indicate that \textbf{enabling users to customize their personal XAI needs is crucial in practice}.

\subsubsection{\textbf{Same feedback can be resolved with different AI explanations.}}

Additionally, we observe that the same writing feedback can be resolved with different AI explanations. 
As shown in Table~\ref{tab:use_pattern}, we demonstrate two use pattern examples to resolve each type of AI prediction feedback as the ``hints'' of how to use XAI within \system systems.

Correspondingly, we also find different users choose different AI explanations to resolve similar problems. For instance, when users receive a suggestion to rewrite the sentence into another aspect label, some participants directly ask for \emph{counterfactual explanations} to change the label (\eg P1, P7, P8), whereas others might refer to \emph{similar examples} to understand the conference published sentences first, and then revise their own writings (\eg P2, P6, P9, P11). Further, even the same people could use different XAIs based on different scenarios. As P1 mentioned \emph{``If time is urgent, I'll use counterfactual explanation because they are straightforward. However, when I have more time, I'll use similar example explanations because I can potentially learn more writing skills from them.''}

\begin{table*}
\includegraphics[width=1\textwidth]{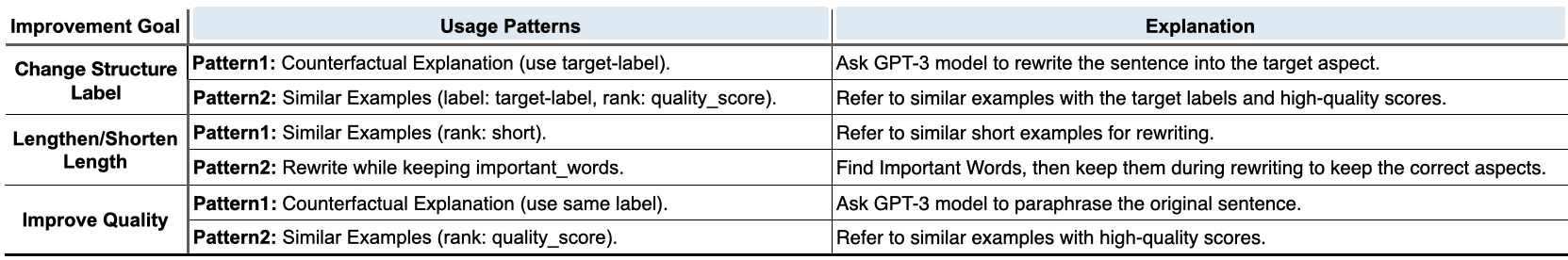}
  \caption{Examples of Use Patterns shown in the ``Tutorial'' explanations suggested by the \system system.}
  \label{tab:use_pattern}
\end{table*}

\section{Discussion and Limitations}
\label{sec:discussion}

In this work, we propose \system as a unified XAI interface in the form of conversations. We especially incorporate practical user demands, representing as the four design principles collected from the formative study, into the \system design.
As a result, users are able to better leverage the multi-faceted, mixed-initiated, context-aware, and customized AI explanations in \system to achieve their tasks (\eg scientific abstract writing). The \system design and findings can potentially shed light on developing more useful XAI systems. 
Additionally, we have released the core codes of unified XAI APIs and the complete code base of \system. The \system can be generalized to a variety of applications since the unified XAI methods and interface are model-agnostic.

In this section, we further elaborate on the core ingredients for useful XAIs based on user study observations with \system, the system generalizability, and the empirical limitations.
As a novel model of a unified XAI interface using conversations, we believe it provides a valuable grounding on how future conversational XAI systems should be developed to better meet real-world user demands.

\subsection{Crucial Ingredients of Useful XAI}

We design \system system as a prototypical yet potential solution of \textbf{useful AI explanation systems} in real-world tasks. 
The rationale is to mitigate the gaps between the practical, diverse, and dynamic user demands of existing AI explanations via a unified XAI interface in the form of conversations. 
Especially, we aim to probe ``\emph{what are the crucial ingredients of useful XAI systems?}'' during the one formative study and two human evaluation tasks. 
In summary, we elaborate on our preliminary findings of useful XAI systems should potentially incorporate four factors, 
including:
`` \textbf{Integrated XAI interface}
+ \textbf{proactive XAI tutorial}
+ \textbf{customized XAIs}
+ \textbf{lightweight XAI display}''.
We elaborate on each ingredient with supportive evidence in our studies for more details.
\paragraph{\emph{\textbf{Integrated XAI interface accessible to multi-faceted XAIs.}}}
In Sec~\ref{sec:use-pattern} and Figure~\ref{fig:user_needs}, we demonstrate diverse XAI user needs and usage patterns from empirical observations. This indicates that XAI user demands are generally dynamically changed across different users and over time. 
Therefore, it is essential to empower users to choose the appropriate XAIs on their own preferences. 
Users can therefore leverage an integrated XAI interface with access to multi-faceted XAIs for their needs.

\paragraph{\emph{\textbf{Proactive XAI usage tutorial.}}}
From the formative study, we learned that it is difficult for users to figure out ``how to leverage and combine the power of different XAI types to finish their practical goals''. This finding motivates the ``Mix-initiated'' design principle, and resulting in designing XAI ``tutorial'' explanations to instruct users.
Moreover, the two user studies provide evidence (\ie in Sec~\ref{sec:xai_tutorial} and Figure~\ref{fig:user_needs}) that users indeed request many XAI tutorial explanations during the writing tasks, but the requested amount is gradually decreased as the users getting more proficient in using the \system system.

\paragraph{\emph{\textbf{Customized XAI interactions.}}}
Users commonly demand more controllability in generating AI explanations. We observed these user demands from both the formative study (\ie leading to ``Controllability'' design principle), and two user studies. More quantitatively, we provide evidence (in Sec~\ref{sec:xai_customization} and Figure~\ref{fig:user_needs}) that although only 3 out of 10 XAI types allow customization, all users leverage XAI customization to generate XAIs. Further, the demands for XAI customization increase over time.

\paragraph{\emph{\textbf{Lightweight XAI display with details-on-demands.}}}
By conducting user studies with both \system and \baseline, we observe that users prefer the XAI interface to be versatile yet simple. Regarding this, a details-on-demand approach using conversations (\eg \system) is more appropriate, as the users can directly pinpoint the expected XAI type as they need. We provide supportive evidence by comparing \system (details-on-demand) and \baseline (full initial disclosure) in Sec~\ref{sec:task1} and Sec~\ref{sec:task2}.

\subsection{Limitations}
\label{sec:limitation}

Although the \system performs mostly better in assisting users in understanding the writing feedback and improving their scientific writings, 
there are still factors and limitations to be noted when deploying \system in practice. 
Here, we discuss potential obstacles they faced and potential fixes to improve \system.

\textbf{Users have a steeper learning curve to use \system.}
In interviewing the users about the advantages and disadvantages of the two systems, we found participants, especially those with less AI knowledge, experienced a steeper learning curve to use the \system -- That says, participants need more effort to learn what answers they can expect from the XAI agent. 
In comparison, they think \baseline is much simpler to interact with because all the answers they can get are displayed in the interface.
However, some participants also mentioned that they would like to spend the efforts to learn \system since it provides more potential explanations to be used. 
From the above observation, we deem that the \system system can be improved by providing the ``instruction of system capability range'' at the initial user interaction stages, and this learning effort will disappear when users interact more with \system in the long run.

\textbf{The performance of writing models and XAI algorithms influence the user experience of \system.}
Another phenomenon we observed is that the under-performed model and XAI algorithm quality can influence the user experience, such as trust and satisfaction. 
Note that in real-world AI tasks, humans are commonly motivated to use the XAI methods to analyze AI predictions, such as improving writing performance according to the AI writing feedback with the help of \system's XAI methods. However, there are situations that AI writing feedback is misaligned with human judgment. In these situations, users commonly ignore the misaligned feedback which can potentially reduce satisfaction and trust in the AI prediction models. 
To mitigate this issue, we posit two actions to resolve: i) it is important to align the AI models' predictions and feedback with human judgment before asking users to leverage analysis methods (\eg XAIs in \system) to explain or interact with the AI predictions.  ii) if the AI task is difficult thus, it's inevitable to occur misalignment (\eg the scientific writing task in this study), enabling human intervention in the models' prediction outputs can alleviate the harm to user experience. For example, when P4 met the misalignment between the model output and his own judgment, he mentioned, ``it would be great if I can manually make the model ignore this review so that the score can reflect my performance more fairly.''

\subsection{Future Directions}
\label{sec:future}

\paragraph{\emph{\textbf{Contextualize for the right user group.}}}
%
During the studies, we found different users with different backgrounds requested diverse levels of AI explanation details for the same XAI question. For instance,  when asking for the model description explanations, AI experts mostly looked for more model details such as the model architecture, how it was trained, etc. In contrast, participants less familiar with AI knowledge only wanted to see the high-level model information, such as who released the model and if it is reliable, etc. 
The observation echoes the motivating example used by \citet{nobani2021towards}, and indicates that users who have different backgrounds need different granularity levels of AI explanations.
While most XAI methods tend to provide user-agnostic information, it might be promising to wrap them based on intended user groups, \eg with non-experts getting the simplified versions with all the jargon removed or explained.
Prior work has also noted that users' perceptions on automated systems can be shaped by conceptual metaphors~\citep{khadpe2020conceptual}, which is also an interesting presentation method to explore.


\paragraph{\emph{\textbf{Characterize the paths and connections between XAI methods.}}}
We observe two interesting usage patterns of XAI methods in \system: 
First, different XAI methods can serve different roles in a conversation. 
For example, explanations on the training data information and model accuracy are static enough that it is sufficient to only describe them once in the \system tutorial; feature attributions and model performance confidence tend to be treated as the basic explanation and initial exploration points, whereas counterfactual explanations are most suitable for follow-ups 
Second, some explanation methods can lead to natural drill-downs. 
For example, we may naturally consider editing the most important words to get counterfactual explanations, \emph{after} we identify those words in feature attributions).
If we more rigorously inspect the best roles of, and links between, explanation methods, we may be able to create a graph connecting them. 
Tracing the graph should help us understand and implement what context should be kept for what potential follow-ups.

Meanwhile, while we encourage continuous conversations, we also observe that as the conversation becomes longer, the earlier information is usually flushed out, and it becomes hard to stay on top of the entire session.
Some users suggested promising directions, one participant recommended ``slicing the dialogue into sessions, where each session only discusses one specific sentence.''
Alternatively, advanced visual signals that reflect conversation structures~\citep{jurafsky2000speech} (\eg the hierarchical dropdown in Wikum reflecting information flow~\citep{zhang2017wikum}) could help people trace back to earlier snippets.

\paragraph{\emph{\textbf{Incorporate multi-modality.}}}
While our current controls and user queries tend to be explicit, prior work envisioned much more implicit control signals. 
For example, \citet{lakkaraju2022rethinking} envisioned the Natural Language Understanding unit should be able to parse sentences like ``Wow, it’s surprising that...'', decipher users' intent on querying outlier feature importance, and provide appropriate responses. Identifying users' emotional responses to certain explanations (\eg surprised, frustrated, affirmed) could be an interesting way to point to potential control responses.

Though natural language interaction is intuitive, not all information needs to be conveyed through dialog. 
Inspired by \baseline's flat learning curve, a combination of natural language inquiry and traditional WIMP interaction could make the system easier to grasp. Future work can survey how people might react to buttons or sliders that allow them to control the number of words or the number of similar examples to inspect.

\section{Conclusion}
\label{sec:conclusion}

In this study, we present \system, a system to support scientific writing via conversational AI explanations. Informed by linguistic properties of human conversation and empirical formative studies, we identify four design principles of Conversational XAI. That says -- these systems should address various user questions (``multi-faceted''), provide details on-demand (``controllability''), and should actively suggest and accept follow-up questions (``mix-initiative'' and ``context-aware drill-down'').
We further build up an interactive prototype to instantiate these rationales, in which paper writers can interact with various state-of-the-art explanations through a typical chatbot interface.
Through 21 user studies, we show that conversational XAI is promising for prompting users to think through what questions they want to ask, and for addressing diverse questions.
We conclude by discussing the use patterns of \system, as well as implications for future conversational XAI systems.

\begin{acks}
We thank Ruchi Panchanadikar for her amazing help in optimizing the UI visualization and functions, Yuxin Deng for her thoughtful comments on improving the user studies, and Reuben Lee for his helpful work on improving the UI details. We also thank all the users for participating in the formative and formal studies, and providing insightful feedback. We thank the reviewers for their constructive feedback.
\end{acks}

\label{sec:ack}

\bibliographystyle{ACM-Reference-Format}
\bibliography{papers}

\newpage
\appendix

\section{Appendix}
\label{sec:appendix}

\begin{table*}
  \includegraphics[width=\textwidth]{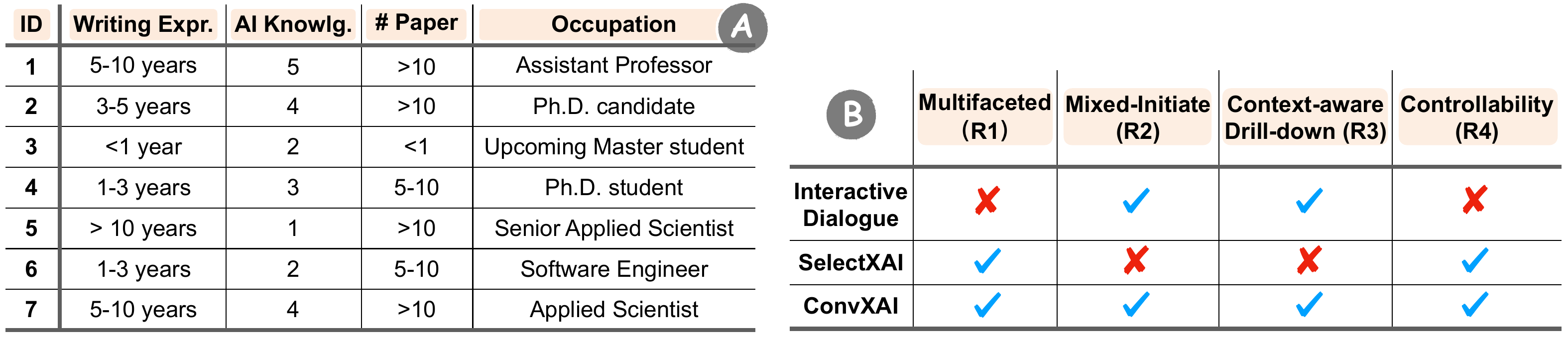}
  \vspace{-0.em}
  \caption{(A) The demographic statistics of the users in the formative study. We recruit seven participants with diverse backgrounds and occupations in order to capture the user needs for the conversational XAI system in more comprehensive views. (B) The four design principles for conversational XAI systems summarized from the formative study. We further compare the existing systems (\emph{i.e.,} Interactive Dialogue~\citep{tsai2021exploring,sun2022exploring}), the baseline (\emph{i.e.,} SelectXAI) and our proposed \system system, regarding these four principles.}
  \label{fig:formative_table}
\end{table*}

\begin{table*}
  \includegraphics[width=1.0\textwidth]{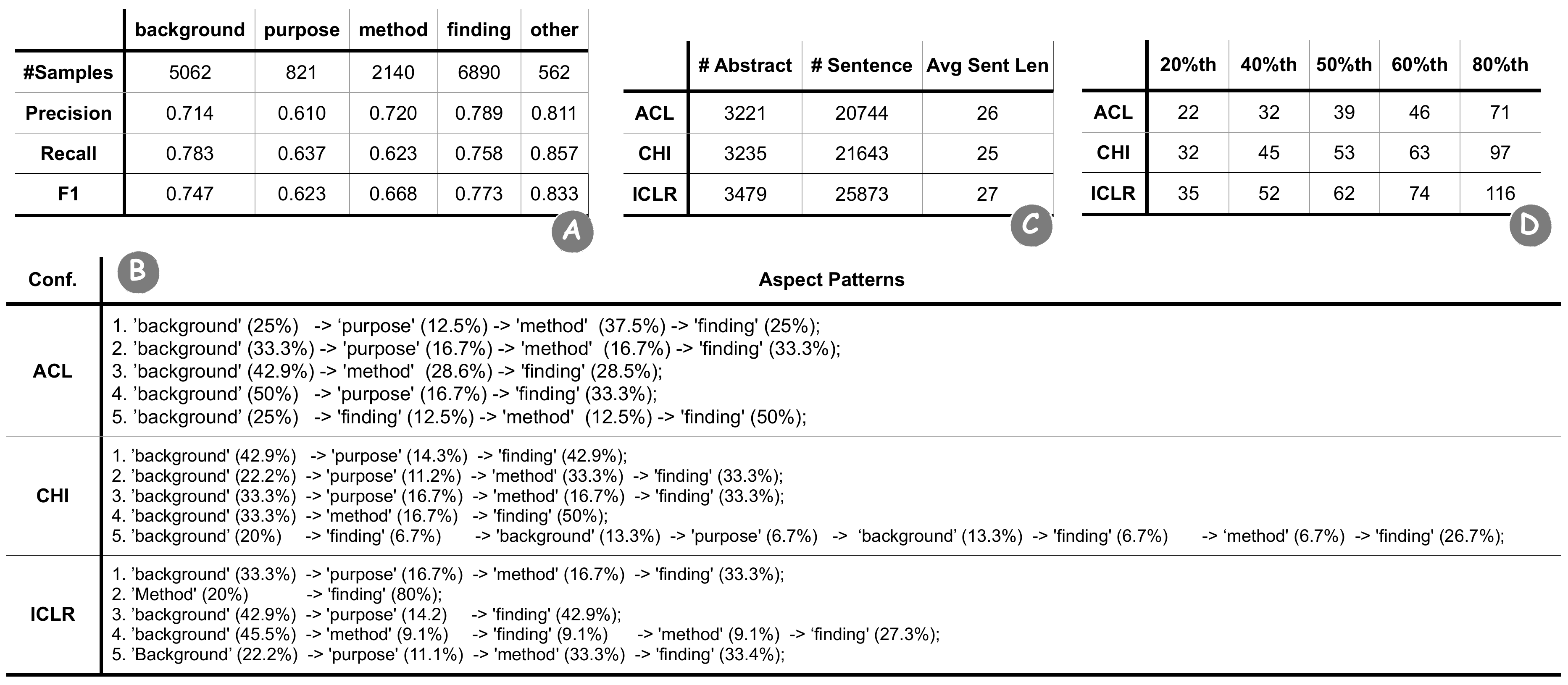}
  \vspace{-0.em}
  \caption{The summary of writing models' performance. The writing structure model performance (with fine-tuned Sci-BERT language model) is shown in (A); (B) shows the extracted five aspect patterns for each conference; the data statistics of three conferences in terms of abstract number, sentence number and average sentence length in (C) and the quality score distribution in (D).
  }
  \label{fig:model_performance}
\end{table*}

\subsection{Formative Study}
\label{appdx:formative}

\subsubsection{Participants Details.} 
In order to capture the user demands of conversational XAI systems from more comprehensive and representative views, we recruited seven participants with diverse backgrounds and occupations in the formative study. The demographic statistics of the seven participants are summarized in Table~\ref{fig:formative_table}(A).
Specifically, 
we invited 7 participants, including 3 females and 4 males. 
In detail, we collected and recorded the participants' information according to the criteria as:
\textbf{Writing Expr.} (\emph{i.e.,} how many years of scientific writing experience do they have?): <1 year; 1-3 years; 3-5 years; 5-10 years; >10 years; 
\textbf{AI Knowlg.} (\emph{i.e.,} what level of AI Knowledgeability would they describe themselves?): 5 - I am a machine learning expert;  4 - I know a lot about machine learning;  3 - I know some knowledge about machine learning; 2 - I know little knowledge about machine learning; 1 - I never heard about machine learning.
\textbf{\# Paper} (\emph{i.e.,} how many submitted papers do they have?): <1;  1-3; 3-5; 5-10; >10.
\textbf{Occupation}: we also record the occupation of each participant.

\subsection{Writing Model Performance}
\label{appdx:model_performance}

We summarize the writing model performance in the Figure~\ref{fig:model_performance}. We can observe the writing structure model performance of the fine-tuned Sci-BERT language model is shown in Figure~\ref{fig:model_performance}A. The model accuracy is 0.7453. Figure~\ref{fig:model_performance}B shows the extracted five aspect patterns for each conference. Further, we can see the data statistics of three conferences in terms of abstract number, sentence number and average sentence length in Figure~\ref{fig:model_performance}C and the quality score distribution in Figure~\ref{fig:model_performance}D.




\end{document}